\documentclass[a4paper,12pt]{article}
\usepackage{jheppub,booktabs}
\usepackage{amsmath,amssymb}
\usepackage{xcolor,colortbl}
\usepackage{cancel}
\usepackage{mathrsfs}
\usepackage{graphicx}
\usepackage{hepunits} 
\usepackage{enumerate}
\usepackage[utf8x]{inputenc}
\usepackage{slashed}

\usepackage{physics} 
\usepackage{subcaption} 

\newcommand{\zP}{z^{\prime}}

\preprint{MITP-25-085}

\author[a]{Andrei Angelescu,} 
\author[a]{Andreas Bally,}
\author[a]{Florian Goertz,}
\author[b]{Sascha Weber}

\affiliation[a]{Max-Planck-Institut f{\"u}r Kernphysik\\ Saupfercheckweg 1, 69117 Heidelberg, Germany}
\affiliation[b]{PRISMA$^+$ Cluster of Excellence \& Mainz Institute for Theoretical Physics, FB 08 - Physics, Mathematics and Computer Science, Johannes Gutenberg-Universität Mainz, Staudingerweg 9, 550099 Mainz, Germany}

\emailAdd{florian.goertz@mpi-hd.mpg.de}
\emailAdd{wesascha@uni-mainz.de}

\title{Gauge Coupling Unification in Gauge-Higgs GUT: Theory and Phenomenology}

\abstract{We present a concise survey of the running of gauge couplings in realistic models of gauge-Higgs grand unification in a slice of AdS$_5$ space and investigate their potential unification.
Besides unifying the gauge groups of the Standard Model, these models can address various unresolved puzzles, such as the lightness of the Higgs boson and the strong hierarchies within fermion masses and mixings, as well as provide a common origin of the gauge symmetries and the sector that spontaneously breaks them. At the same time, they furnish interesting LHC signatures in the form of TeV-scale resonances of the $X,Y$--like bosons, providing a trace of the grand-unified group, accessible at low energies.

Using the method of Planck-brane correlators allows us to evolve the couplings consistently from the electroweak scale up to the Planck scale, avoiding shortcomings of other frequently-used approaches and including the effects of bulk scalars, fermions, and gauge-bosons within a common framework. We thereby revisit, contrast, and supplement results in the literature, the latter for example by including brane masses and the gauge-Higgs vacuum expectation value. Moreover, in a phenomenology section, we apply our results to the concrete case of Georgi--Glashow--like unification with a ${\rm SU(6)} \supset {\rm SU(5)}$ symmetry in the 5D bulk, presenting a quantitative survey of the quality of unification. We find that grand unification is possible in such models in the presence of moderately large brane kinetic terms.}

\date{\today}

\begin{document}
\maketitle

\section{Introduction}
There is a long history in physics in trying to unify apparently different phenomena in a single theoretical description, seeking a common origin. Prominent examples are grand unified theories (GUTs)~\cite{Georgi:1974sy,Pati:1974yy}, that integrate the known gauge forces into a simple gauge group. A further step would be to also include the spontaneous breaking of the corresponding gauge symmetries. This can be achieved in models with extra dimensions, allowing to embed the Higgs sector within a higher-dimensional gauge field, via the idea of gauge-Higgs grand unified theories (GHGUTs)~\cite{Hall:2001zb,Nomura:2001mf,Burdman:2002se,Haba:2004qf,Lim:2007jv,Hosotani:2015hoa,Furui:2016owe,Hosotani:2016njs,Hosotani:2017edv,Hosotani:2017hmu,Maru:2019lit,Maru:2019bjr,Funatsu:2019xwr,Englert:2019xhz,Englert:2020eep,Angelescu:2021nbp,Maru:2022hex,Maru:2022mbi,Angelescu:2022obm,Hamaguchi:2022yjo,Angelescu:2023usv,Maru:2024ljf,Cacciapaglia:2024duu,Komori:2025wji}, thereby also solving the hierarchy problem of GUTs~\cite{Gildener:1976ai,Weinberg:1978ym,Wells:2025hur} (see also \cite{Agashe:2005vg,Frigerio:2011zg,Barnard:2014tla} for 4D-dual {\it composite Higgs} GUTs). One of the most attractive features of GUTs is the unification of gauge couplings: the values of the gauge couplings at a low scale have a common origin at a high scale and the relation between the two is given by the {\it running} of the couplings, determined by the renormalization group equations (RGEs). In this work we build upon existing techniques and extend them to study explicitly the running of gauge couplings in a realistic GHGUT, also comparing to different approaches used in the literature. 

As a specific example we take a minimal viable model based on a SU(6) symmetry in the bulk of a warped extra dimension, which includes a SU(5) grand unified group, presented in \cite{Angelescu:2021nbp,Angelescu:2022obm}, but our results can easily be applied to other models, too. In the model of \cite{Angelescu:2021nbp,Angelescu:2022obm}, common challenges in GHGUT constructions are overcome by directly breaking to the Standard Model (SM) gauge group at one boundary of the (warped) extra dimension, which allowed for the first time to correctly reproduce all fermion masses from 5D spinors together with generating a natural Higgs mass $m_h \approx 125$ GeV and avoiding ultra-light exotic particles.\footnote{It is interesting to note that minimal GHGUTs also allow for a group-theoretical cancellation mechanism for fermionic corrections to the Higgs boson mass~\cite{Angelescu:2023usv}.} Here we will investigate if the same model also allows for successful unification of gauge couplings at a high scale.

Calculating the running of the couplings in the usual 4D space--time and the application to 4D GUTs is well known, a prime example of this is the original Georgi--Glashow model based on the GUT group SU(5) \cite{Georgi:1974sy}, although the gauge couplings do not meet precisely enough given our current low energy limits on them.\footnote{This can, for example, be improved in a supersymmetric set--up.} To see if gauge coupling unification occurs in the GHGUT models based on warped extra dimensions, one has to know how gauge couplings evolve in $\mathrm{AdS}_5$, which comes with additional challenges compared to the 4D case.

For flat extra dimensions the RGE has been calculated e.g. in~\cite{Dienes:1998vh,Dienes:1998vg}, finding a power--law running, in contrast to the logarithmic running in 4D. Given masses of the lowest KK excitations of the order of \si{\TeV} this would necessitate unification at the \si{\TeV} scale.
As has been pointed out in~\cite{Pomarol:2000hp}, this is no longer true in warped extra dimensions: there can be a logarithmic evolution of gauge coupling up to the Planck scale even with Kaluza-Klein (KK) excitations at the \si{\TeV} scale, the latter possibly providing experimental hints for the existence of the GUT group. 

Subsequently, the running and the application to GUTs in warped spaces have been further studied in~\cite{Agashe:2002bx,Agashe:2002pr,Contino:2002kc,Choi:2002wx,Choi:2002zi,Choi:2002ps,Gherghetta:2004sq,Agashe:2005vg,Frigerio:2011zg,Randall:2001gc,Randall:2001gb,Randall:2002qr,Falkowski:2002cm,Goldberger:2002cz,Goldberger:2002hb,Goldberger:2003mi}, using different approaches and methods. The difference between them can broadly be classified according to how they connect with low energy phenomenology. In~\cite{Pomarol:2000hp,Agashe:2002bx,Agashe:2002pr,Contino:2002kc,Choi:2002wx,Choi:2002zi,Choi:2002ps} the KK picture is used to calculate the zero-mode gauge coupling by an appropriate regularization of the infinite KK tower. The resulting evolution is only valid for momenta below the \si{\TeV} scale, since for higher momenta the zero mode becomes strongly coupled, as has been pointed out e.g.\ in~\cite{Contino:2002kc}. Alternatively, one can also use holographic methods to relate warped extra dimensions to dual 4D strongly coupled theories, i.e., Composite Higgs (CH) theories. In the context of CH theories the running and unification have been studied in~\cite{Agashe:2005vg,Gherghetta:2004sq,Frigerio:2011zg}. Furthermore, one can use deconstruction to find an approximate 4D theory and study the running within it, as has been done in~\cite{Randall:2002qr,Falkowski:2002cm}. Staying in the 5D theory an approach inspired by holography has also been taken up in~\cite{Randall:2001gc,Randall:2001gb} by using a position--dependent cut--off. Given the explicit dependence of the final $\beta$--functions on the regulator, the result is hard to interpret~\cite{Choi:2002ps,Goldberger:2002hb} and, similar to the KK picture, it focuses on the coupling of the zero mode and is only valid for low momenta. Moreover, we found that one encounters additional challenges when one tries to apply this method to fermions in the loop, leading to non-gauge invariant results~\cite{Weber:2022hpi}.

In this work, we employ the Planck–brane correlator approach~\cite{Goldberger:2002cz,Goldberger:2002hb,Goldberger:2003mi}, which avoids the limitations discussed above. Inspired by holography as well, it uses the brane--to--brane correlator on the Planck--brane, a correlation function which is valid for momenta above the \si{\TeV} scale, to derive the gauge--coupling running in the effective theory up to the (high) unification scale. Furthermore, it can be matched at low momenta to the results from the calculations using the zero mode~\cite{Contino:2002kc}, see Sec.~\ref{sec:matching}. 
In particular, we present numerical results for the quality of unification in the minimal SU(6) GHGUT.

This paper is structured as follows: In Section~\ref{sec:overview_RS} we will provide the necessary overview of warped extra dimensions and introduce the SU(6) GHGUT of~\cite{Angelescu:2021nbp}. Section~\ref{sec:overview} gives an overview on unification in GUTs and a quick intuition to understand the differences emerging in (warped) extra dimensions. We will use the Planck--brane correlator to derive the gauge-coupling running for general scalar, fermion, and gauge fields in Section~\ref{sec:derivation}, reproducing and extending results available in the literature. This is followed by a brief discussion on the matching at low energies in Section~\ref{sec:matching}. In Section~\ref{sec:application} we will apply our results to specific realistic incarnations of SU(6) GHGUTs to highlight their differences and comment on the quality of precision unification. Finally, we will summarize our results and conclude in Section~\ref{sec:conclusion}.

\section{Warped Extra Dimensions and SU(6) GHGUT} \label{sec:overview_RS}

In this section we give a brief overview on the details of warped extra dimensions necessary for this work. More comprehensive reviews can be found e.g. in~\cite{Arkani-Hamed:2000ijo,Csaki:2004ay,Csaki:2005vy,Sundrum:2005jf,Contino:2010rs,Gherghetta:2010cj,Ponton:2012bi,Csaki:2015xpj}. Working in conformal coordinates, the metric of the warped AdS$_5$ space is given by
\begin{align}\label{eq:RS_metric}
    \mathrm{d}s^2=\left(\frac{1}{kz}\right)^2(\eta_{\mu\nu}\mathrm{d}x^\mu \mathrm{d}x^\nu -\mathrm{d}z^2) \, ,
\end{align}
where $z\in[1/k,1/T]$, and $k\sim M_{\mathrm{pl}}$ ($T\sim \si{\TeV})$ corresponds to the position of the UV (IR) brane, truncating the extra dimension. Given these energy scales, one also refers to the UV brane as the Planck brane and to the IR brane as the \si{\TeV} brane. We collect the relevant parts of the action for the various fields that propagate in the extra dimension, as well as their propagtors and vertices, in Appendix \ref{sec:Feynman_rules}. Eventually, the full SM will live in the compactified five-dimensional space-time.

There are at least two useful ways to connect these 5D theories with low energy observations. The first is a so called Kaluza-Klein (KK) decomposition, where one decomposes a 5D field $\phi(x,z)$ into a set of 4D fields $\phi_{n}(x)$ as (see, e.g.,~\cite{Sundrum:2005jf,Gherghetta:2010cj,Ponton:2012bi,Csaki:2015xpj})
\begin{align}\label{eq:KK_scalar}
    \phi(x,z)=\sum_n f_{n}(z) \phi_{n}(x) .
\end{align}
Integrating over the extra dimensions and taking into account the KK wavefunctions $f_{n}(z)$, results in a theory of a tower of 4D fields (KK modes) $\phi_{n}(x)$ with increasing mass $m_{n}$. Depending on the boundary conditions (BCs) at $z=1/k,1/T$, there can be a massless zero mode solution ($n=0$) in the spectrum, resembling a known SM-like particle, and all higher massive modes ($n\geq 1$) have masses starting from the KK scale $m_{\mathrm{KK}} \sim T \sim \si{\TeV}$~\cite{Sundrum:2005jf,Gherghetta:2010cj,Ponton:2012bi,Csaki:2015xpj}. Furthermore, the localization of the bulk profile solutions $f_{n}(z)$ along the extra dimension will have important consequences for the running, as we will see later.

The second way to relate to low energy phenomenology is by using holography. Inspired by the AdS/CFT correspondence~\cite{Maldacena:1997re}, one can look at the field values on the UV brane and integrate out the rest of the bulk and the IR brane. By doing this one obtains a weakly coupled sector, the elementary sector, corresponding to fields on the UV brane which are weakly coupled to a strongly interacting sector, the composite sector, corresponding to the fields which have been integrated out~\cite{Arkani-Hamed:2000ijo,Contino:2010rs,Gherghetta:2010cj,Csaki:2015xpj}. This latter sector is described by a conformal field theory for energies below the UV scale with conformal invariance beeing spontaneously broken at the IR scale leading to confinement.

\begin{figure}[!t]
    \centering
    \raisebox{1.5mm}{\includegraphics[width=0.45\textwidth]{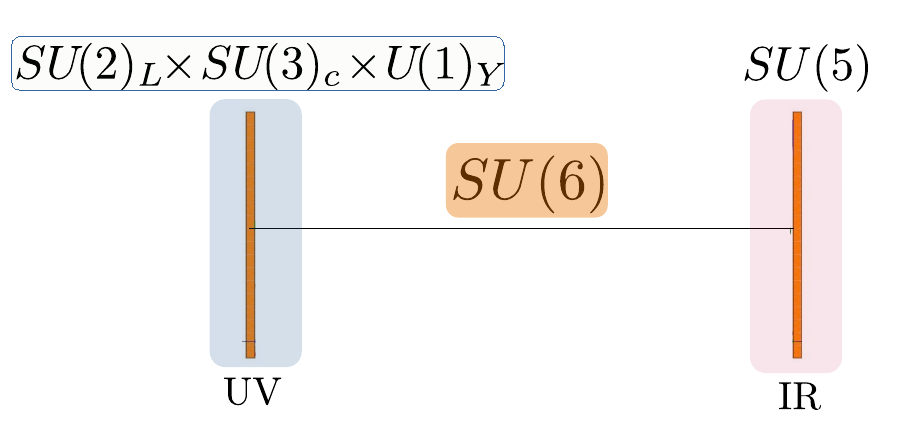}}\qquad
    \includegraphics[width=0.45\textwidth]{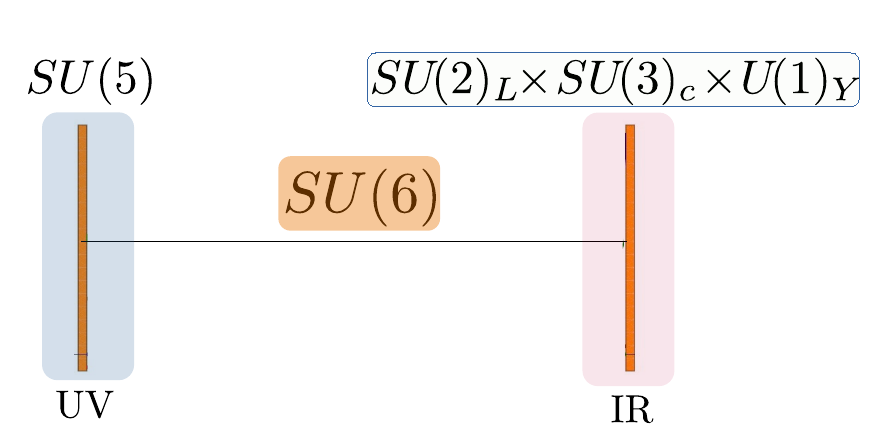}
\caption{Setup of the $G_{\textrm{SM}}^{(\textrm{UV})}$ (left) and $G_{\textrm{SM}}^{(\textrm{IR})}$ (right) models, indicating the unbroken gauge groups on the respective branes. See text for details.}
\label{fig:Setup}
\end{figure}
Fields can be unified in various ways in this extra dimensional setup. A 5D gauge field $A_M^a=(A_{\mu}^a,A_5^a)$ can be split into a 4D vector part $A_{\mu}^a$ and a 4D scalar part $A_5^a$. The idea of gauge-Higgs unification (GHU)~\cite{HOSOTANI1983193,HOSOTANI1983309,MANTON1979141,FAIRLIE197997} is to embed the Higgs field in a zero mode of the 4D scalar $A_5^a$. Indeed, a potential for $A_5^a$ is forbidden by 5D gauge symmetry. At one-loop, upon symmetry breaking at the boundaries, a finite potential is generated possibly inducing a 4D vev $\langle A_{5}^a(x)\rangle=v^a$. A Gauge--Higgs Grand Unified Theory (GHGUT) tries to combine this with the idea of a GUT, where also the gauge interactions are unified into a single gauge field under one gauge group. Thus GHGUTs can lead to a complete description of the full bosonic sector of the SM in terms of one single 5D gauge field of a (simple) gauge group. For simple gauge groups, the minimal group is SU(6) and in~\cite{Angelescu:2021nbp,Angelescu:2022obm} it was shown that a correct radiative Higgs potential as well as the SM flavor structure can be reproduced successfully -- even predicting the correct down-quark hierarchies -- if an appropriate simple breaking pattern is employed. In the following we will describe this set--up, studied below, briefly.

For the gauge sector we employ two incarnations of the model, denoted by $G_{\textrm{SM}}^{(\textrm{UV})}$  ($G_{\textrm{SM}}^{(\textrm{IR})}$) in \cite{Angelescu:2022obm}. Here, the bulk gauge symmetry SU(6) is broken to the SM group by non-universal BCs for the gauge fields on the UV (IR) brane (respecting only $G_{\textrm{SM}}$ and allowing no other vector-boson zero-modes), while on the other brane the unbroken group is SU(5), see Fig.~\ref{fig:Setup}. 
For the $G_{\textrm{SM}}^{(\textrm{UV})}$ setup, the BCs explicitly read~\cite{Angelescu:2022obm}
\begin{equation}\label{eq:BC_UV_model}
\begin{split}
A_\mu&= \left( \begin{array}{cc|ccc|c}
 \textcolor{blue}{(++)} & \textcolor{blue}{(++)} & (-+) & (-+) & (-+) & (--)\\
 \textcolor{blue}{(++)} & \textcolor{blue}{(++)} & (-+) & (-+) & (-+) & (--)\\
 \hline
 (-+) & (-+) & \textcolor{blue}{(++)} & \textcolor{blue}{(++)} & \textcolor{blue}{(++)} & (--)\\
 (-+) & (-+) & \textcolor{blue}{(++)} & \textcolor{blue}{(++)} & \textcolor{blue}{(++)} & (--)\\
 (-+) & (-+) & \textcolor{blue}{(++)} & \textcolor{blue}{(++)} & \textcolor{blue}{(++)} & (--)\\
 \hline
 (--) & (--) & (--) & (--) & (--) & (--)\\
\end{array} \right) \, ,
\end{split}
\end{equation}
where a $+(-)$ denotes a Neuman (Dirichlet) BC on the respective brane and the upper-left (central) highlighted block belongs to $\text{SU(2)}_{\rm L}$ ($\text{SU(3)}_{\rm c}$). Only fields with $(++)$ BCs feature a zero mode. For the case of $G_{\textrm{SM}}^{(\textrm{IR})}$, the UV and IR BCs are interchanged.

Moving on to the fermion content, the minimal viable embedding consists of a $\bf{1}, \bf{6}, \bf{15}$ and a $\bf{20}$, which are the four smallest SU(6) representations \cite{Angelescu:2021nbp}. For $G_{\textrm{SM}}^{(\textrm{UV})}$ the decomposition into $G_{\textrm{SM}}$ representations and associated BCs read~\cite{Angelescu:2022obm}
\begin{align}
\label{eq:IR-embedding}
    {\bf 20} \rightarrow & q^\prime{\bf ( 3,2)}_{1/6}^{+,-}  \oplus {\bf (3^*,1)}_{-2/3}^{+,-} \oplus  e^{c \prime} {\bf (1,1)}_1^{+,-}  \notag \\ 
     & \oplus {\bf (3^*,2)}_{-1/6}^{+,-}  \oplus u {\bf (3,1)}_{2/3}^{-,-} \oplus{\bf (1,1)}_{-1}^{+,-}, \notag \\
    {\bf 15} \rightarrow  & q {\bf (3,2)}_{1/6}^{+,+} \oplus {\bf (3^*,1)}_{-2/3}^{-,+} \oplus e^c {\bf (1,1)}_1^{+,+} \notag \\  
    & \oplus d^\prime {\bf (3,1)}_{-1/3}^{-,+}\oplus  l^{c \prime}{\bf (1,2)}_{1/2}^{-,+}, \notag \\
   {\bf 6} \rightarrow & d  {\bf (3,1)}_{-1/3}^{-,-} \oplus l^c  {\bf (1,2)}_{1/2}^{-,-} \oplus \nu^c {\bf (1,1)}_0^{+,+}, \notag \\
   {\bf 1} \rightarrow &  \nu^{c \prime}{\bf (1,1)}_{0}^{+,-} \, ,
\end{align}
where we present as superscripts the BCs for the left--handed (LH) spinors with those for the right--handed (RH) fields being flipped.
The fields which feature light zero modes are suggestively labeled with the corresponding SM symbols.\footnote{We use $\Psi^c_L\equiv(\Psi^c)_L=(\Psi_R)^c$. Therefore the label L/R indicates the transformation properties under the Lorentz group. For the SM leptons, embedded as conjugates, this means the singlet and doublet electron are denoted by $e^c_L$ and $l^c_R$, respectively.} The fields with a prime are connected to the unprimed ones via brane masses on the IR brane, whose precise structure can be found in~\cite{Angelescu:2022obm}, but will not be important for the results derived here. For $G_{\textrm{SM}}^{(\textrm{IR})}$ the same decomposition applies just with the UV and IR BCs interchanged.

As it will turn out, for $G_{\textrm{SM}}^{(\textrm{UV})}$, an important parameter determining the contribution of fermions to the RGE is the bulk mass parameter $c=m/k$. We use the ranges denoted by $\mathrm{C}1$, as listed in Table~\ref{tab:cs-all}, which can reproduce the SM flavor structure~\cite{Angelescu:2022obm}.
\begin{table}[ht]
    \centering
    \begin{tabular}{rrr|c|c}
        \multicolumn{3}{c|}{C1 \& C2} & C1 & C2 \\
        \hline
        $c_{20,1} < -1/2$, & $c_{20,2} < -1/2$, & & $-1/2 < c_{20,3} < +1/2$ & $c_{20,3} > +1/2$ \\
        $c_{15,1} > +1/2$, & $c_{15,2} > +1/2$, & & $-1/2<c_{15,3} < +1/2$ & $c_{15,3} > +1/2$ \\
        $c_{6,1} < -1/2$, & $c_{6,2} < -1/2$, & $c_{6,3} < -1/2$ &  & 
    \end{tabular}
    \caption{Assignment of fermion bulk parameters $c_i\in[-1,1]$ used in~\cite{Angelescu:2022obm} (C1) compared to a theory resembling the composite Higgs GUT of~\cite{Agashe:2005vg} (C2). See text for details.}
    \label{tab:cs-all}
\end{table}
To compare to other composite Higgs GUT models in the literature, e.g.~\cite{Agashe:2005vg}, we will also use the assignment $\mathrm{C}2$, also listed in Table~\ref{tab:cs-all}. The precise numerical values of these bulk masses will not be important for the results of this work, but the qualitative difference between these assignments becomes clear when looking at their 4D composite duals. For $\mathrm{C}2$, all SM fermion fields are elementary except for the RH top $t_R$, which is composite, as is often assumed in CH theories. In the case of $\mathrm{C}1$, the first and second generations of SM fermions are elementary and one has a non--elementary third generation with the RH top $t_R$, the third generation LH quark doublet $q_3$ and the RH tau lepton $\tau_R$.

\section{Gauge-Coupling Evolution in 5D}\label{sec:overview}

In flat 4D theories the action of a gauge field can be written as
\begin{align}\label{eq:4D-gauge-action}
    S \!\supset\! \int \mathrm{d}^4 x \bigg(\!-\frac{1}{4g^2} F_{\mu\nu} F^{\mu\nu} \!\bigg) ,
\end{align}
where we have rescaled the originally canonical normalized fields by $A_{\mu} \to g^{-1} A_{\mu}$, such that the 4D gauge coupling $g$ only appears as the coefficient of the gauge kinetic term and thus only in the propagator of the gauge field. Regularization and renormalization leads to a scale dependence of the gauge coupling. If the model is described by a GUT, like the Georgi--Glashow (GG) model based on SU(5)~\cite{Georgi:1974sy}, the gauge couplings should meet, up to threshold effects, at some energy scale. This can also be explored by studying the differential running of the three SM couplings, i.e., the running of $\Delta\alpha_{i1}=\alpha_i-\alpha_1$, with the fine structure constants $\alpha_i=g_i^2/(4\pi)$. Alternatively, assuming unification, the meeting point is given by only two couplings, say $\alpha_{1},\alpha_{2}$, from which the low energy value of the third, $\alpha_{3}$, can be \textit{postdicted}. For the GG model one obtains $\delta_{3}=(\alpha_3^{\text{theo}}-\alpha_3^{\text{exp}})/\alpha_3^{\text{exp}} \approx 40\%$, for which the discrepancy with the measured value is too large, although the situation improves when adding additional fields, like in supersymmetric theories. That the gauge couplings evolve logarithmically is no coincidence and can already be anticipated from the mass dimension of the operator coefficients in the action. The divergence structure has the same form as \eqref{eq:4D-gauge-action}, thus can be renormalized via the gauge coupling $g$. Since we have $[g^{-2}]=0$ for the mass dimension of the gauge coupling, the dependence on a regularization mass scale $\Lambda$ can only be logarithmically with $\log(\Lambda)$. 

This situation changes considerably in higher dimensional theories. For a slice of $\mathrm{AdS}_5$ the gauge field part of the action reads
\begin{align}\label{eq:action}
    S \!\supset\! \int \mathrm{d}^4 x \mathrm{d} z \left(\frac{1}{kz}\right) \bigg(\!-\frac{1}{4g_5^2} F_{MN} F^{MN} + \frac{\lambda_{k}}{4}F_{\mu\nu} F^{\mu\nu} \delta\!\left(z-\frac{1}{k}\right) + \frac{\lambda_{T}}{4}F_{\mu\nu} F^{\mu\nu} \delta\!\left(z-\frac{1}{T}\right) \!\bigg) ,
\end{align}
where we have again normalized the gauge fields such that the bulk gauge coupling $g_5$ appears in front of the kinetic term and not in the covariant derivative. Here, $[g_5^{-2}] = 1$ and there will be power-law divergent contributions to the gauge boson self energy scaling as $\Lambda^1$ with $\Lambda$ an UV cutoff~\cite{Goldberger:2002hb}. Additionally, because of the finite extra dimension, there are brane kinetic terms with dimensionless coefficients $[\lambda_k]=[\lambda_T]=0$ which can be used to renormalize logarithmic divergencies, i.e. $\log(\Lambda)$~\cite{Goldberger:2002hb}. The power-law divergencies dominate for energies above the inverse size $L^{-1}$ of the extra dimension, at which point the effective description of the theory breaks down~\cite{Goldberger:2002cz,Goldberger:2002hb,Goldberger:2003mi}. Below these energies, the logarithmic divergencies dominate instead. If we assume KK excitations with masses of the $\order{\si{\TeV}}$ this implies for flat extra dimensions a size of $\order{\si{\TeV}^{-1}}$ but, due to the warping, in warped extra dimensions typically a size of $\order{M_{\mathrm{Pl}}^{-1}}$. Thus for warped extra dimensions we can have logarithmic running of gauge couplings up to the Planck scale if we can use a well defined correlation function to access it. The correlation function of zero mode fields will not be calculable for momenta above the $\si{\TeV}$ scale as the zero mode becomes strongly coupled~\cite{Contino:2002kc}.  Instead, as mentioned before, we will use the Planck--brane correlator~\cite{Goldberger:2002cz,Goldberger:2002hb,Goldberger:2003mi}, inspired by holography, which is valid up to the energies of the order $k\sim M_{\mathrm{Pl}}$ and, as we will see in the next section, has the right divergence structure to be renormalized by the brane coupling $\lambda_k$ on the Planck brane.

\section{Explicit Derivation of the Running}\label{sec:derivation}

We now turn to the explicit analysis of the Planck--brane correlator, i.e. the 5D gauge field propagator whose endpoints lie on the Planck brane. We initially closely follow \cite{Goldberger:2003mi}, but including fermion brane masses in Section \ref{sec:running_fermions}, and add our own calculation for gauge fields in Section \ref{sec:running_gauge_fields}. 
Since we rescaled the fields in Eq.~\eqref{eq:action} such that the bulk gauge coupling $g_5$ appears in front of the kinetic term, the two-point correlator will be proportional to $g_5^2$ at tree level. In line with this, we define the general effective gauge coupling of the elementary sector $g(p)$ via the Planck--brane correlator\footnote{\label{fn:GI} Strictly speaking, the propagator is not gauge invariant for non--abelian gauge fields in contrast to the Wilson loop \cite{Goldberger:2003mi}. Although we work with this gauge-dependent definition, gauge invariance implies the equivalence for different gauge choices.} as
\begin{equation}\label{eq:def_gauge_coupling}
   \Delta_{\mu\nu}(q^2) = \int \mathrm{d}^4x \, e^{iq\cdot x} \expval{A_{\mu}\left(x,\frac{1}{k}\right)A_{\nu}\left(0,\frac{1}{k}\right)} = \frac{g^2(q)}{q^2} \eta_{\mu \nu} + \text{gauge dep.\, .}
\end{equation}
We will drop the gauge dependent parts in the following and focus only on the terms proportional to $\eta_{\mu \nu}$, which will be sufficient for calculating the running. Furthermore, we will employ dimensional regularization using $d=4-2\epsilon$ and work directly in Euclidean $\text{AdS}_{d+1}$ space to make the resulting integrals well defined. 

First, we look at the tree--level value of this coupling. Using the propagators\footnote{For convenience, we work in the following with the euclidean action and, because of our rescaling $A_M \to g_5^{-1}A_M$, the propagator includes an extra factor of $g_5^2$.} given in Appendix \ref{sec:prop_gauge_boson} we have
\begin{equation}
    \frac{g^2(q)}{q^2}\delta_{\mu\nu}\bigg\rvert_{\text{tree}} =\Delta_{\mu\nu}(q^2)\bigg\rvert_{\text{tree}} =  g_5^2 G_{1,q}\left(\frac{1}{k},\frac{1}{k}\right) \delta_{\mu\nu}\ ,
\end{equation}
where $G_{1,q}\left(z,\zP\right) = G_{1}\left(q,z,\zP\right)$ is the transverse component of the gauge boson 5D propagator with 4D (euclidean) momentum $q$ from position $\zP$ to $z$ (see Eq.~\eqref{eq:gauge_prop_tensor_reduction}). It is instructive to expand this for energies $q \ll T$ and $T \ll q \ll k$, where
\begin{equation}
    g^2(q)\big\rvert_{\text{tree}} \simeq \begin{cases}
        g_5^2 \frac{k}{\ln(\frac{k}{T})} & q \ll T \\
        g_5^2 \frac{k}{\ln(2e^{-\gamma_E}\frac{k}{q})} & T \ll q \ll k
    \end{cases} \, . \label{eq:tree_gauge_coupling}
\end{equation}
Notice that for $q \ll T$ this agrees with the tree--level value for the zero mode coupling
\begin{equation}
    g^2(q \ll T)\big\rvert_{\text{tree}} \simeq g_5^2 \frac{1}{L} \equiv g_{4D,0}^2 \, ,
\end{equation}
with $L = \frac{1}{k} \ln(\frac{k}{T})$ the separation between the UV and IR branes. The tree--level coupling of the zero mode $g_{4D,0}$ depends on the large logarithm $\ln(\frac{k}{T})$ as if it would be a running coupling which has been run down from the Planck scale $k$ to the low scale $T$. This is commonly referred to as "tree--level running" \cite{Pomarol:2000hp,Arkani-Hamed:2000ijo,Randall:2001gb,Goldberger:2002hb,Goldberger:2003mi}. In fact, in the holographic dual this value arises from the universal leading pure CFT correction \cite{Goldberger:2002hb}. Note that for $q\gg T$ the effective gauge coupling $g(q)$ depends already on the momentum $q$ at tree--level.

\begin{figure}
    \centering
    \includegraphics[width=0.3\linewidth]{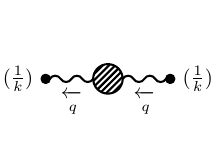}
    \caption{Planck-brane correlator determining the evolution of the gauge coupling at one loop.}
    \label{fig:gauge_boson_blob}
\end{figure}

Our goal is now to look at loop corrections to the Planck--brane correlator, depicted in Fig.~\ref{fig:gauge_boson_blob}. In general, this takes the form
\begin{equation}
    \Delta_{\mu\nu}(q^2)\big\rvert_{\text{1--loop}} = \int \mathrm{d} z \mathrm{d} \zP g_5^2 G_{1,q}\left(\frac{1}{k},z\right) \times \Pi_{\mu\nu,q}(z,\zP) \times g_5^2 G_{1,q}\left(\zP,\frac{1}{k}\right) \, ,
\end{equation}
where $z,\zP$ can lie in the bulk. Let us look at the external propagators more closely. Since we are interested in the coupling of the gauge fields which feature zero modes, i.e. the SM gauge fields, we can take the 5D propagators from Appendix~\ref{sec:prop_gauge_boson} using $(+,+)$ BCs (i.e.\ $P_{UV}=0,P_{IR}=0$) and no UV brane masses (i.e\ $M_{UV}^2=0$). We are interested in the region $q\gg T$ for which we expand the propagator
\begin{equation}
    G_{1,q} (\frac 1 k,z) \approx \left( kz \right)^{(d-3)/2} \, \frac{1}{q} \, \left(\sqrt{kz}\frac{K_{\frac{d}{2}-1}(q z)}{K_{\frac{d}{2}-2}(\frac q k)}\right) \, \Omega(z)^\dagger\,,
\end{equation}
where $\Omega(z)$ parametrizes the effects of EWSB (see Eq.~\eqref{eq:goldstone-matrix-gauge}).
Due to 
\begin{equation}\label{eq:Bessel_fun_expansion}
    K_{\frac{d}{2}-1}(q z)\propto \begin{cases}
        \sqrt{\frac{\pi}{2 q z}}e^{-qz} & \text{for } qz > 1 \\
        (q z)^{-(\frac{d}{2}-1)} & \text{for } qz < 1
    \end{cases} ,
\end{equation}
the propagator is exponentially suppressed for $qz\gg1$ and the bulk of the contribution comes from the smallest $z$ values, thus near the Planck--brane.  This implies $\Omega(z) \to \Omega(1/k)=1$ as well as that the dynamics on the IR brane are completely irrelevant.\footnote{This limit is equivalent to taking $1/T \to \infty$. Moreover, the BCs and brane masses on the IR brane do not affect the behavior of the propagators near the UV brane and for convenience we choose in the following $P_{IR}=0,M_{IR}=0,\Omega(z)=1$ for all fields.}

It will turn out that for the momentum range $T \ll q \ll k$ and for the divergence structure relevant for the differential running (further contributions will be neglected in the following), the $z$--integration and the integration over loop momenta will separate for the case of scalars, fermions and gauge bosons in the loop. In this case, the integration over $z$ can be performed explicitly leading always to two copies of the tree--level Planck--brane propagator. Explicitly, we find
\begin{equation}\label{eq:1_loop_PB_correlator}
    \Delta_{\mu\nu}(q^2)\big\rvert_{\text{1--loop}} \simeq g_5^2 G_{1,q}\left(\frac{1}{k},\frac{1}{k}\right) \times \Pi_{\mu\nu}(q^2) \times g_5^2 G_{1,q}\left(\frac{1}{k},\frac{1}{k}\right) \, ,
\end{equation}
where $\Pi_{\mu\nu}(q^2)\simeq q^2 \delta_{\mu\nu}\Pi(q^2)$ turns out to be equal to the gauge boson self energy in $4$--dimensional flat space \cite{Goldberger:2002hb,Goldberger:2003mi}. In particular for scalar fields this can be intuitively understood, since there is effectively only one KK-mode which is not suppressed at the Planck-brane \cite{Goldberger:2003mi}. For other fields the result Eq.~\eqref{eq:1_loop_PB_correlator} will be true as well, thus to calculate the running it will be sufficient to know the results of $\Pi(q^2)$, which we will explicitly calculate in the following sections for the different fields. The divergent term $\Pi(q^2)$ can be regularized in the usual way, e.g. via dimensional regularization, leading to $1/\epsilon$ poles as well as logarithms of the form
\begin{equation}\label{eq:4D-self-energy}
    \Pi(q^2) = \frac{b}{16\pi^2}\left[\frac{1}{\epsilon}-\gamma_E+\log(4\pi)+\log(\frac{\mu^2}{q^2})+\dots\right]\,,
\end{equation}
with the coefficient $b$ depending on the fields in the loop, see the upcoming subsections for the explicit values. Notice that Eq.~\eqref{eq:1_loop_PB_correlator} can be interpreted as a vertex $\Pi_{\mu\nu}(q^2)$ localized on the Planck brane. 

For small enough $\lambda_k$ the brane kinetic term localized on the Planck brane can be treated perturbatively and gives (dropping again gauge dependent terms)
\begin{equation}
\label{eq:BKT}
    \Delta_{\mu\nu}(q^2)\big\rvert_{\text{brane}} = g_5^2 G_{1,q}\left(\frac{1}{k},\frac{1}{k}\right) \times (-\lambda_k q^2 \delta_{\mu\nu}) \times g_5^2 G_{1,q}\left(\frac{1}{k},\frac{1}{k}\right) \, .
\end{equation}
This has the same structure as the divergent loop contribution, as it has to be to renormalize the divergence in \eqref{eq:4D-self-energy} via the coupling $\lambda_k \to \lambda_k(\mu)$, where $\mu$ is the subtraction scale. Adding this to the tree-level coupling we find, after the usual resummation, the running coupling 
\begin{align}
\label{eq:runcoup}
    \frac{1}{g^2(q^2)}\bigg\rvert_{\text{tree}+\text{1--loop}} &= \frac{1}{g^2(q^2)}\bigg\rvert_{\text{tree}} + \lambda_k(\mu) - \frac{b}{16\pi^2} \log(\frac{\mu^2}{q^2})
\end{align}
In the next subsections we will clarify the details for the different fields with different BCs, brane masses, and Gauge--Higgs vev.

\subsection{Scalar Fields}\label{sec:running_scalar}

Given that there are no fundamental scalars in our theory, we just use them as an explicit straightforward example to demonstrate the formalism. The following calculation is essentially the same as in \cite{Goldberger:2003mi}, so we will only highlight the important steps in the calculation and its interpretation, with a focus on the application to the GHGUT. We consider a $(+,+)$ complex scalar field with no brane mass ($P_{UV} = 0,M_{UV}^2=0$), with a finite bulk mass ($m\neq 0)$, coupled to a $U(1)$ gauge field. From the last section we saw that the Gauge-Higgs vev can be neglected for $q\gg T$ (see discussion under Eq.~\eqref{eq:Bessel_fun_expansion}).

\begin{figure}
\centering
\includegraphics[width=0.65\linewidth]{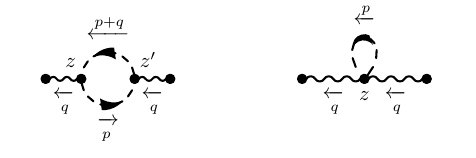}
\caption{Diagrams contributing to the RGE for internal scalars.}
\label{fig:scalar_loops}
\end{figure}

There are two diagrams contributing, which are depicted in Fig.~\ref{fig:scalar_loops}, with extra factors of $(R/z)^{d-1}$ from the warped metric compared to scalar QED. We obtain $\Pi_{\mu\nu,q}(z,\zP)=\Pi_{\mu\nu,q}^{(1)}(z,\zP)+\Pi_{\mu\nu,q}^{(2)}(z,\zP)$, with
\begin{gather}
    \Pi_{\mu\nu,q}^{(1)}(z,\zP) = \int_p \left[\left(\frac{1}{kz}\right)^{d-1}(2p+q)_{\mu}\right]G_{0,p}(z,\zP) G_{0,p+q}(\zP,z)\left[\left(\frac{1}{k\zP}\right)^{d-1}(2p+q)_{\nu}\right] \ , \notag \\
    \Pi_{\mu\nu,q}^{(2)}(z,\zP) = \delta(z-\zP) \int_p \left[\left(\frac{1}{kz}\right)^{d-1}2\delta_{\mu\nu}\right]G_{0,p}(z,z) \ ,
\end{gather}
where $G_{0,p}(z,\zP) = G_{0}(p,z,\zP)$ is the 5D scalar propagator (see Eq.~\eqref{eq:scalar-prop-diff-eq}) and we neglected the longuitudinal, i.e., gauge dependent part of the gauge boson propagator as described in the last section. The factors in square brackets describe the vertices, which are the same as in scalar QED but dressed with factors coming from the warped metric. 

For $q\ll k$, loop momenta above the scale $k$ cannot give rise to non--analytic effects \cite{Goldberger:2003mi}, thus we expand the internal propagators in $T\ll p,(p+q)\ll k$. As the leading contribution from the $z$--integration comes from the region near $z=1/k$ we can also expand the internal propagators in $pz,(p+q)z,p\zP,(p+q)\zP \ll 1$ to determine the leading non--analytic behavior. In the following we will only focus on these non--analytic parts, since they determine the divergence and therefore the running of the couplings. The analytical parts only change the constants appearing in Eq.~\eqref{eq:4D-self-energy}, and dropping them is an implicit choice of scheme.\footnote{Note that such a scheme choice does not influence the low energy predictions for physical observables if used consistently \cite{Goldberger:2003mi}.} Including also an expansion in $m /k \ll d$, the internal propagators, derived in Appendix~\ref{sec:prop_scalar}, can be written as
\begin{gather}\label{eq:scalar_prop_expanded}
    G_{0,p}(z,\zP) \simeq k\left(2-d\right)\left(kz\right)^{\frac{d-1}{2}} \left(k\zP\right)^{\frac{d-1}{2}} \left(kz\right)^{-\alpha} \left(k\zP\right)^{-\alpha} \left(\frac{1}{p^2+m_d^2}\right) \ ,
\end{gather}
where we introduced $m_d^2\equiv\frac{-2+d}{d}m^2$ and $\alpha=(\sqrt{d^2+4m^2/k^2}-1)/2 \simeq (d-1)/2$. We see that the functional dependence on $z$ and $p$ has split into separate parts allowing us to split the corresponding integrals. 

The integration over $z$ can be performed together with the external propagators in the limit $k\gg q$ and gives for both $i=1,2$ 
\begin{equation}
    \Delta_{\mu\nu}^{(i)}(q^2)\big\rvert_{\text{1--loop}} \simeq \left[g_5^2G_{1,q}\left(\frac{1}{k},\frac{1}{k}\right)\right]^2 \Pi_{\mu\nu}^{(i)}(q^2)\,,
\end{equation}
where the remaining momentum integrals read
\begin{gather}
    \Pi_{\mu\nu}^{(1)}(q^2) = \int_p \left[(2p+q)_{\mu}\right]\left[(2p+q)_{\nu}\right] \frac{1}{p^2+m_d^2}\frac{1}{(p+q)^2+m_d^2}\notag \\
    \Pi_{\mu\nu}^{(2)}(q^2) = \int_p \left[2\delta_{\mu\nu}\right]\frac{1}{p^2+m_d^2}\,.
\end{gather}
These are the standard integrals from scalar QED in $d$--dimensional euclidean space which can be evaluated, as in 4D, around $d=4-2\epsilon$. Adding them, and taking the limit $m\to0$ for simplicity, we obtain the general result of equations \eqref{eq:1_loop_PB_correlator} and \eqref{eq:4D-self-energy} with $b=-1/3$ and therefore, after renormalizing with the Planck-brane kinetic term \eqref{eq:BKT}, the running is given by equation \eqref{eq:runcoup} with the same $b$.
This can be generalized for the case of SU(N) by including the appropriate group theoretic factors.
In total, we can summarize that the contribution of a scalar field with $(+)$ BCs on the Planck brane to the $\beta$--function is that of the corresponding 4D scalar field with a mass $m_d$.

If we change now the BCs of the scalar field to $(-)$ BCs ($P_{UV} = 1$) one gets a similar structure as \eqref{eq:scalar_prop_expanded} but with a mass $m_{d}^2\approx2(-2+d)k^2$, so a mass of the order of the Planck scale. These terms only give contributions to the difference in gauge couplings for $q\gtrsim k$. For the energies we are interested in ($q\ll k$) this means that a field with $(-)$ BCs on the UV brane does not contribute to the differential running at all. A similar conclusion will apply also for fermions and gauge bosons in the loop and will be very important in deriving the running in the GHGUT.

Lastly, there is also an intuitive understanding of the above from the KK picture. For $(-)$ BCs the individual fields vanish on the Planck brane, giving no contribution to the non-analytic part of the Planck--brane correlator. For $(+)$ BCs most modes will be localized towards the IR brane, apart from one single mode, which is the zero mode in the case of $m=0$ and a mode with a KK mass of the order $m$ for $m\neq 0$. This intuition will carry over to the other spin fields.

\subsection{Fermion Fields}\label{sec:running_fermions}

Next we consider Fermion fields. After discussing the important changes from the scalar case for vanishing brane masses, we will also consider the effect of non-zero fermion brane masses to be able to apply the results to the model of~\cite{Angelescu:2021nbp,Angelescu:2022obm}.

\begin{figure}
\centering
\includegraphics[width=0.3\linewidth]{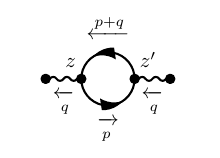}
\caption{Diagram contributing to the RGE for internal fermions.}
\label{fig:fermion_loops}
\end{figure}

There is the usual QED type of diagram for the self energy (c.f.~Fig.~\ref{fig:fermion_loops}), leading to
\begin{gather}\label{eq:fermion_loop_equation}
    \Pi_{\mu\nu,q}(z,\zP) = \int_p \Tr[\left(\frac{1}{kz}\right)^{d}\gamma_{\mu} G_{1/2,p}(z,\zP) \left(\frac{1}{k\zP}\right)^{d}\gamma_{\nu} G_{1/2,p+q}(\zP,z)] \ .
\end{gather}
Here $G_{1/2,p}(z,\zP) = G_{1/2}(p,z,\zP)$ is the fermion propagator (see Eq.~\eqref{eq:ferm-prop-diff-eq}) and the vertices are like the ones in standard QED but dressed with factors from the warped metric. By splitting the propagators into the different chirality components (see \eqref{eq:ferm-prop-dirac-basis}) the Dirac structure can be simplified. Focusing first on $(+)$ BCs on the UV brane, we find that only the $\Tilde{G}_{LL}$ part in \eqref{eq:ferm-prop-dirac-basis} will give rise to relevant, non-analytic structures. The precise form depends on the value of the bulk mass parameter $c$. For $c>\frac{1}{2}$ we find after the expansion, as described in Section~\ref{sec:running_scalar}
\begin{align}\label{eq:fermion_prop_expansion}
    G_{1/2,p} (z,\zP) &\simeq \left( kz \right)^{d/2} \left( k\zP \right)^{d/2} \, P_L i  \slashed{p} \, \tilde{G}_{LL} \notag \\
    &\simeq \left( kz \right)^{d/2} \left( k\zP \right)^{d/2} \,k(-1+2c)\left( kz \right)^{-c} \left( k\zP \right)^{-c} \ \left(\frac{ P_L i  \slashed{p}}{p^2}\right)\,.
\end{align}
Again, the prefactor, when integrated over $z,\zP$ with the external propagators, leads to $\left[g_5^2 G_{1,q}(1/k,1/k)\right]^2$ and the remaining momentum integral is reminiscent of the usual 4D integral of QED with a massless fermion. The corresponding contribution to the $\beta$--function will thus be given as in Eq.~\eqref{eq:runcoup} with $b=-2/3$ corresponding to a Weyl fermion, or to a Dirac Fermion including an additional factor $1/2$ coming from $P_L$, reflecting that only the LH component contributes for the latter. The result can easily be extended to non-abelian gauge fields by multiplying $b$ with the Dynkin index $T(R_f)$ of the corresponding fermion representation $R_f$. Contrary to the scalar case the bulk mass parameter $c$ does not give a mass term to this propagator, but changes the localization, and thus the divergence structure. 

This can be seen by studying the case $c<\frac{1}{2}$, for which we find
\begin{align}
    G_{1/2,p} (z,\zP) &\simeq \left( kz \right)^{d/2} \left( k \zP \right)^{d/2} \, P_L i  \slashed{p} \, \tilde{G}_{LL} \notag \\
    &\simeq \left( kz \right)^{d/2} \left( k\zP \right)^{d/2} 4^c \frac{\Gamma(\frac{1}{2}+c)}{\Gamma(\frac{1}{2}-c)}\left( kz \right)^{-c} \left( k\zP \right)^{-c} \ \left(\frac{ P_L i  \slashed{p} }{k(p/k)^{1+2c}}\right)\,.
\end{align}
For $-\frac{1}{2}<c<\frac{1}{2}$ no large logarithms arise since the propagator is more singular in the UV. We note that UV power corrections will be universal if the symmetry is broken only by boundary effects (like in the present case), and therefore will not contribute to the differential running~\cite{Goldberger:2003mi}.
For $c<-\frac{1}{2}$ the propagator is analytic with no contribution to the differential running, too. Lastly, we can talk about the case $c=\pm\frac{1}{2}$ by taking the limit $c\to\pm\frac{1}{2}$ after the integration over $z$ finding the same momentum structure as for $c>\frac{1}{2}$ ($c<-\frac{1}{2}$) for $c\to+\frac{1}{2}$ ($c\to-\frac{1}{2}$). Thus for $(+)$ BC on the UV brane only in the case $c\geq \frac{1}{2}$ there is a contribution to the running equal to half the contribution of Dirac fermions in 4D flat space.
For $(-)$ BCs on the UV brane one finds a similar result except now only the RH part contributes for $c\leq-\frac{1}{2}$. This is not surprising as the equation for the fermion propagator are symmetric under simultaneously changing $(+)\leftrightarrow(-),c\leftrightarrow-c,L\leftrightarrow R$. 

Having established the general behavior of fermionic contributions, we study now the case where they are connected via brane masses, which has not been considered in the literature before. For simplicity we focus on two flavors $\Psi=(\psi_1 \ \psi_2)^T$ with opposite BCs $(+)$ and $(-)$, respectively. They can be connected via a UV brane mass $m_{UV}$.\footnote{Here we write the $2 \times 2$ matrix in flavor space $M_{UV} = m_{UV}\begin{pmatrix}
    0 & 0 \\
    1  & 0
\end{pmatrix}$, such that $m_{UV}$ is a scalar.} This is already quite involved since we need to consider the propagator for each of the four flavor components and each of the four chirality components. Moreover, in each of these cases the pole structure will look differently, depending on the possible assignments of the bulk mass parameters $c_1,c_2$. For example, for $c_1>\frac{1}{2}$ and $c_2 < - \frac{1}{2}$ and writing $\expval{\psi_i \Bar{\psi_j}} \sim G_{1/2}^{ij}$ for the propagator we find 
\begin{align}
    G_{1/2,LL,p}^{11} (z,\zP) &\simeq -\left( kz \right)^{d/2} \left( k\zP \right)^{d/2} \,k(1-2c_1)\left( kz \right)^{-c_1} \left( k\zP \right)^{-c_1} \nonumber \\
    &\quad \times \left(1+\abs{m_{UV}}^2\frac{1+2c_2}{1+2c_1}\right)\left(\frac{ P_L i  \slashed{p}}{p^2 + \abs{m_{UV}}^2k^2(1-2c_1)(1+2c_2)}\right)\,, \\
    G_{1/2,RR,p}^{22} (z,\zP) &\simeq -\left( kz \right)^{d/2} \left( k\zP \right)^{d/2} \,k(1+2c_2)\left( kz \right)^{c_2} \left( k\zP \right)^{c_2} \nonumber \\
    &\quad \times \left(1+\abs{m_{UV}}^2\frac{1-2c_1}{1-2c_2}\right)\left(\frac{ P_R i  \slashed{p}}{p^2 + \abs{m_{UV}}^2k^2(1-2c_1)(1+2c_2)}\right)\,.
\end{align}
Without the brane mass ($m_{UV}=0$) this case corresponds to two Weyl fermions contributing to the running, a left-handed mode for $\psi_1$ ($(+)$ BC and $c_1>1/2$) and a right-handed mode $\psi_2$ ($(-)$ BC and $c_2<-1/2$). Including brane masses on the UV brane, these modes combine to form a Dirac fermion with mass $m^2 = \abs{m_{UV}}^2k^{2}(1-2c_1)(1+2c_2)$. Their combination would only contribute to the differential running above this mass scale. However, in the GHGUT model incarnations considered in this work, we never connect such fermion fields via Planck brane masses. In these cases we get a contribution as without a brane mass. 

For example, for the same assignments of BCs and now $c_1>\frac{1}{2}$ and $-\frac{1}{2}<c_2<\frac{1}{2}$ we obtain instead
\begin{align}
    G_{1/2,LL,p}^{11} (z,\zP) &\simeq \left( kz \right)^{d/2} \left( k\zP \right)^{d/2} \,k(-1+2c_1)\left( kz \right)^{-c_1} \left( k\zP \right)^{-c_1} \ \left(\frac{ P_L i  \slashed{p}}{p^2}\right)\,, \\
    G_{1/2,RR,p}^{22} (z,\zP) &\simeq \left( kz \right)^{d/2} \left( k\zP \right)^{d/2} \,4^{-c_2}\frac{\Gamma(\frac{1}{2}-c_2)}{\Gamma(\frac{1}{2}+c_2)}\left( kz \right)^{c_2} \left( k\zP \right)^{c_2} \left(1+\abs{m_{UV}}^2\frac{1-2c_1}{1-2c_2}\right) \nonumber \\
    &\quad \times \left(\frac{ P_R i  \slashed{p}/k}{ (p/k)^{1-2c_2}+\abs{m_{UV}}^2(1-2c_1)(1+2c_2)2^{-1-2c_2}\frac{\Gamma(-\frac{1}{2}-c_2)}{\Gamma(\frac{1}{2}+c_2)}}\right)\,,
\end{align}
where only the LH part of $\psi_1$ contributes to the running as in the case without the brane mass. We see that for two fermion fields, where only one of them would contribute to the running without the UV-brane mass, the UV brane mass is irrelevant for the contribution to the running. This is the case for all brane masses in the models considered in the following sections.

\subsection{Gauge Fields}\label{sec:running_gauge_fields}

For non-abelian SU(N) gauge fields Ref.~\cite{Goldberger:2003mi} used the Wilson loop to define the effective gauge coupling and then calculated the loop contribution with the help of AdS/CFT arguments. Instead, here we continue to use Eq.~\eqref{eq:def_gauge_coupling} and just calculate loop corrections to the Planck-brane correlator. While strictly speaking this is not gauge invariant, the running of the extracted coupling respects gauge invariance and is physical (see footnote \ref{fn:GI}). Our results are qualitatively consistent with the statements made in \cite{Goldberger:2003mi}.

\begin{figure}
\centering
\includegraphics[width=0.55\linewidth]{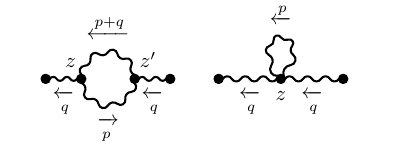}
\caption{Diagrams contributing to the RGE for internal $A_{\mu}$.}
\label{fig:loops_gauge}
\end{figure}

In the following we consider the 4D vector $A_{\mu}$ and scalar $A_5$ contributions of the 5D gauge field separately. For only $A_{\mu}$'s running in the loop, we find a similar contribution as for 4D gauge fields in QCD with the two terms (c.f.~Fig.~\ref{fig:loops_gauge})
\begin{gather}
    \Pi_{\mu\nu,q}^{(1),ab}(z,\zP) = \int_p \left[\left(\frac{1}{kz}\right)^{d-3}P_{\lambda_1\lambda_2\mu}^{a_1 a_2 a}(p+q,-q,-p)\right]G_{1,p,\lambda_1\rho_1}(z,\zP) G_{1,p+q,\lambda_2\rho_2}(z,\zP) \notag \\
    \times \left[\left(\frac{1}{k\zP}\right)^{d-3}P_{\rho_1\rho_2\nu}^{a_1 a_2 b}(-p-q,q,p)\right] \notag \\
    \Pi_{\mu\nu,q}^{(2)}(z,\zP) = \delta(z-\zP) \int_p \left[\left(\frac{1}{kz}\right)^{d-3}N_{\mu\lambda_1\lambda_2\nu}^{accb}\right]G_{1,p,\lambda_1\lambda_2}(z,z)\,,
\end{gather}
with the usual 4D ``three-gluon'' vertex $P_{\lambda_1\lambda_2\lambda_3}^{a_1 a_2 a_3}(p_1,p_2,p_3)$ and 4D ``four-gluon'' vertex, $N_{\lambda_1\lambda_2\lambda_3\lambda_4}^{a_1a_2a_3a_4}$ dressed by factors from the warped metric in the square brackets. In this notation we define $P_{\lambda_1\lambda_2\lambda_3}^{a_1 a_2 a_3}(p_1,p_2,p_3)$ such that all momenta $p_i$ are incoming, $a_i$ are the gauge group indices and $\lambda_i$ the Lorentz indices. 

Focusing first on the case without brane masses, we can use the same approximations as before (sub-Planckian momenta and UV localization) to find that there exists a pole in the internal propagator only for $(+)$ BC where the propagator is explicitly given as
\begin{equation}
    G_{1,p}(z,\zP)\simeq-2\left(\frac{d}{2}-2\right)k \frac{1}{p^2}\,, \label{eq:gauge_boson_prop_expansion}
\end{equation}
While in the limit $d\to4$ this propagator seems to vanish, the prefactor in Eq.~\eqref{eq:gauge_boson_prop_expansion}, after integration over $z,\zP$ and including the external propagators,  leads again to $\left[g_5^2 G_{1,q}(1/k,1/k)\right]^2$ which remains finite in the limit $d\to4$. Note that the ghost propagator (see Eq.~\eqref{eq:ghost_prop_solution}) is described by the same propagator as the gauge fields up to gauge dependent terms. We are then once more left with the standard 4D diagrams which lead to the usual gauge field contribution of a SU(N) gauge field, i.e., $b=11N/3$. 

Next we consider the field $A_5$. As an example, one finds for $(-)$ BC, which corresponds to the case of non-zero $A_5$ on the UV brane and a zero mode in the spectrum, the following propagator expansion for sub-Planckian momenta and UV localization
\begin{equation}
    G_{5,p}(z,\zP)\simeq-\left(kz\right)^{-d+4}\left(k\zP\right)^{-d+4}2^{-5+d} \left(\frac{p}{k}\right)^{5-d} \frac{\Gamma(\frac{d}{2}-2)}{\Gamma(3-\frac{d}{2})}\,.
\end{equation}
This propagator has no momentum pole around $d=4$, and loops with only $A_5$ will not contribute to the running. For diagrams with a $A_{\mu}$ and $A_5$ in the loop, only the propagator of $A_{\mu}$ can have a pole, but we can always shift the momentum variable to make this pole scaleless. Therefore, these diagrams do not give a contribution in dimensional regularization. The same will be true no matter the BCs of $A_5$, leading to the conclusion that none of the scalar parts of gauge fields contribute to the running.

\subsection{Summary of the Contributions}\label{sec:summary_running}

Here we summarize the above results. For all cases, the BCs on the IR brane as well as brane masses on the IR brane and the Higgs vev do not influence if and how a field contributes to the Planck--brane correlator. A 5D scalar contributes to the running only for $(+)$ BCs on the UV brane with a $\beta$--function of a corresponding 4D scalar, with a mass $m_d^2=\frac{-2+d}{d}m^2$ given by the bulk mass $m$. The contribution of a 5D fermion depends on the bulk mass parameter $c$. For $c\geq1/2$ only for $(+)$ BCs there is a contribution to the $\beta$--function equal to a corresponding 4D LH massless fermion. For $-1/2<c<1/2$ there is no contribution to the $\beta$--function, no matter the BCs. For $c\leq-1/2$ only for $(-)$ BCs there is a contribution to the $\beta$--function equal to a corresponding 4D RH massless fermion. The inclusion of UV--brane masses, which connect different fermion fields, does not affect the contribution to the $\beta$--function unless a field with $(+)$ BC and $c_1\geq1/2$ is connected to a field with $(-)$ BC and $c_2\leq-1/2$, in which case these combine giving a contribution to the $\beta$--function like a 4D Dirac fermion with mass $m^2=(1-2c_1)(1+2c_2)\abs{m_{UV}}^2k^2$. Finally, the vector part $A_{\mu}$ of a 5D gauge field $A_{M}=(A_{\mu},A_{5})$ contributes to the running only for $(+)$ BCs on the UV brane with a $\beta$--function of a corresponding 4D gauge field. The scalar part $A_{5}$ of a 5D gauge field $A_{M}=(A_{\mu},A_{5})$ does not contribute to the running for all values of the BCs on the UV brane.

\section{Matching at the Low Scale}\label{sec:matching}

After having outlined the general contributions of fields to the running, we detail in this section how to relate to the measured coupling at low energies. The remaining symmetry on the UV brane determines the contribution of the individual fields as well as which brane kinetic terms are present. For the breaking pattern $G_{\textrm{SM}}^{(\textrm{UV})}$, which breaks on the UV brane to the SM, there are three UV-brane kinetics corresponding to the three gauge fields with zero modes, which we identify with the three gauge fields of the SM. Explicitly we have at one loop for the fine structure constants $\alpha_{i}$ 
\begin{align}
    \alpha_i^{-1}(q^2) &= \alpha_{\text{tree}}^{-1}(q^2) + 4\pi \lambda_{k,i}(\mu) - \frac{b_i}{4\pi} \log(\frac{\mu^2}{q^2})
\end{align}
Note that the uncalculable tree--level contribution is the same for each gauge field and it drops out of the differential running $\Delta\alpha_{i1}^{-1}\equiv\alpha_{i}^{-1}-\alpha_{1}^{-1}$,
\begin{align}
    \Delta\alpha_{i1}^{-1}(q^2) &= 4\pi \Delta\lambda_{k,i1}(\mu) - \frac{\Delta b_{i1}}{4\pi} \log(\frac{\mu^2}{q^2})
\end{align}
where we defined $\Delta\lambda_{k,i1}\equiv\lambda_{k,i}-\lambda_{k,1}$ and $\Delta b_{i1} \equiv b_{i}- b_{1}$. The coefficients $\Delta b_{i1}$ are fully determined by the field content, UV BCs and Lagrangian parameters (e.g. bulk mass parameters). However, one needs further assumptions on the couplings $\lambda_{k,i}$. Since these couplings correspond to brane kinetic terms for the gauge fields, one can think of them as UV-brane gauge couplings $\lambda_{k,i}\sim1/g_{{\rm UV},i}^2$~\cite{Hall:2001pg,Choi:2002ps}. Under the assumption of strong coupling one expects from naive dimensional analysis (NDA) $\lambda_{k,i}(\mu=k)\approx 1/(16\pi^2)$ \cite{Hall:2001pg}, for which these corrections would be small, but without knowledge of a concrete UV completion also larger values seem possible. We will comment more on this issue and the unification of the gauge couplings for the concrete model in the next section. 

The above formula is only valid for momenta above the \si{\TeV} scale, at which it could be matched to results from other methods like \cite{Contino:2002kc}, which focus on the zero mode propagator, if the running below the \si{\TeV} scale should be included. However, as the main contribution to the running comes from the evolution from the Planck scale to the \si{\TeV} scale, we ignore the exact matching and use the usual SM running for momenta below the IR scale $T$.

Before presenting the results for the $G_{\textrm{SM}}^{(\textrm{UV})}$ setup, we briefly discuss the other breaking pattern $G_{\textrm{SM}}^{(\textrm{IR})}$ for the SU(6) GHGUT of \cite{Angelescu:2021nbp,Angelescu:2022obm}, see Section~\ref{sec:overview_RS}, where one breaks to the SM on the IR brane and to SU(5) on the UV brane. In this case, the gauge kinetics $\lambda_{k,i}$ as well as the coefficients $b_i$ will be SU(5) universal, implying that a unification of the gauge couplings would have to happen at the \si{\TeV} scale. However, the values of the measured gauge couplings at low energies are too different to unify at this scale with only logarithmic running.

\section{Application to SU(6) GHGUT}\label{sec:application}

\begin{figure}[th]
\begin{subfigure}[b]{0.5\textwidth}
    \includegraphics[width=\textwidth]{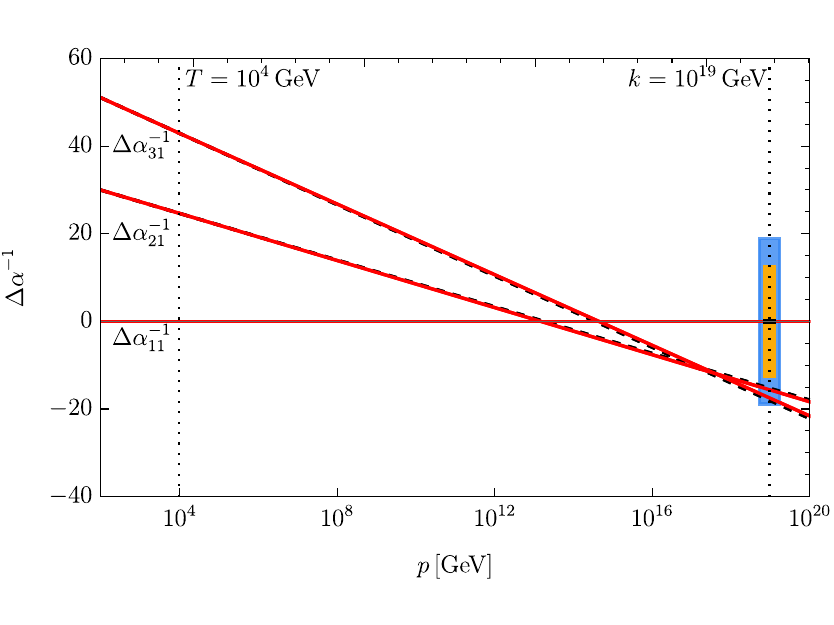}
\end{subfigure}
\hfill
\begin{subfigure}[b]{0.5\textwidth}
    \includegraphics[width=\textwidth]{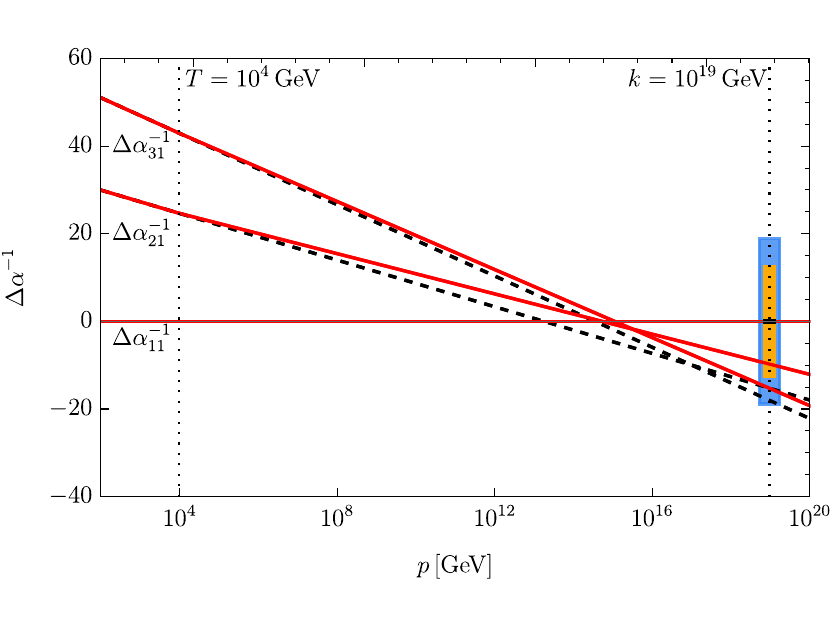}
\end{subfigure}
\caption{Differential running for the fermion assignment C1 (left) and assignment C2 (right) in red, compared to the SM differential running in dashed. The vertical dotted lines display the Planck scale $k$ and the \si{\TeV} scale $T$ below which the SM running is employed. The impact of $\Delta\lambda_{k,i1}$ on the Planck-brane is illustrated by colored bands. The NDA estimate $\Delta\lambda_{k,i1}\approx1/(16\pi^2)$, shown in black, is too small to have a visible vertical extend. We additionally display regions corresponding to moderately large $\Delta\lambda_{k,i1}\lesssim1$ ($\Delta\lambda_{k,i1}\lesssim1.5$) in orange (blue). In the presence of such terms, values of the couplings within these regions allow for a successful embedding in the unified group. See text for details.}
\label{fig:differential_running}
\end{figure}

In the following, we present our numerical results for the $G_{\textrm{SM}}^{(\textrm{UV})}$ model of Section~\ref{sec:overview_RS}. Given our choice of BCs presented in Eqs.~\eqref{eq:BC_UV_model} and \eqref{eq:IR-embedding}, one can directly determine the $\beta$--functions from the bosonic and fermionic field content and the bulk mass parameter assignments C1 and C2 of Table~\ref{tab:cs-all} using the results of Section~\ref{sec:summary_running}. We plot the differential running for both cases in Figure~\ref{fig:differential_running} and will describe the underlying contributions below.
We can conveniently summarize the relevant contributions for each case by looking at the changes compared to the SM. Starting with the bosonic fields, the gauge fields $A_{\mu}$ have $(+)$ BCs on the UV brane only for the SM fields resulting in a contribution to the differential running as in the SM. On the other hand, the Higgs, being embedded in the $A_5$ component, always drops out. Since the contribution of the Higgs to the running of the SM gauge coupling is minuscule, this is a small effect. The contributions of the fermions are much more important.

For C1, the bulk mass parameters of the first two fermion generations are chosen in such a way that only the fields corresponding to the SM fermions are contributing, whereas the primed fields and exotics from Eq.~\eqref{eq:IR-embedding} do not. This is different for the third generation. Here, the $c$ values for the $\bf{20}$ and $\bf{15}$ lie in between $-1/2 < c < +1/2$, so none of the fields is contributing. Thus, compared to the SM, the third generation $u,q$ and $e^c$ do not contribute. Note that they form a complete SU(5) multiplet, which has no effect on the differential running. Hence, the differential running is the same as in the SM, up to the small difference due to the missing Higgs contribution. For C2 the situation changes. Still the first two generations contribute as in the SM, but now also the third generation $q$ and $e^c$ in the $\bf{15}$ contribute as well. Although the $c$ value of the $\bf{20}$ changes too, the third generation $u=t_R$ still does not contribute, but instead the exotics in the second row of the $\bf{20}$ in \eqref{eq:IR-embedding} now contribute to the differential running. Note that they complete the $t_R$ to a SU(5) multiplet and for the differential running one can describe their contribution as a missing $t_R^c$. This is exactly the same scenario for the differntial running as in \cite{Agashe:2005vg}. In the dual theory, one can interpret the assignment C2 as having elementary fields for all SM fields except for the $t_R$ which is composite. Consequently, we need additional exotics to fill out a complete multiplet of the strong sector. Thus, the right panel of Figure~\ref{fig:differential_running} exactly matches Figure 1 in \cite{Agashe:2005vg}.

We can also comment on the unification of the gauge couplings. Due to the presence of the possibly different UV brane gauge kinetic couplings $\lambda_{k,i}$ for each gauge coupling $g_i$, the gauge couplings, strictly speaking, never unify. Nonetheless, if the differences $\Delta\lambda_{k,i}$ are small, one can give a range in which the the low energy gauge couplings should lie, given the presence of a grand unified gauge group in the bulk, and confront it with experimental results. Note that in this model, the GUT breaking scale has to be the Planck--scale, as compared to regular flat 5D SU(5) unification where the unification scale typically would lie around the \si{\TeV} scale. For this we show in Figure \ref{fig:differential_running} a range of possible variations of $\Delta\lambda_{k,i}$. For the assignment C1 one needs at least $\Delta\lambda_{k,i}\approx 1.4$ in order to reproduce the correct low energy couplings, and although the meeting of the gauge couplings for the assignment C2 is much better at the scale $q \sim \SI{e15}{\giga\electronvolt}$, at the scale $q\sim k$ one needs $\Delta\lambda_{k,i}\approx1.2$. If one interprets $\lambda_{k,i}=1/{g_{UV,i}^2}$ one can reproduce this with couplings $g_{UV,i} \sim \mathcal{O}(1)$. 

\section{Conclusions}\label{sec:conclusion}

In this article, we provided a comprehensive condensed survey of the RGE of the gauge couplings in realistic models of gauge-Higgs grand unification, including analytical and numerical results. Such models furnish well motivated extensions of the SM that can address several of its shortcomings, while unifying fundamental concepts in various facets and leading to interesting predictions around the TeV scale (see, e.g., \cite{Angelescu:2022obm,Angelescu:2023usv}). After introducing GHGUTs and specifically a phenomenologically viable variant of Georgi--Glashow--like unification with a ${\rm SU(6)} \supset {\rm SU(5)}$ symmetry in the bulk, we presented general results for the running of the gauge couplings in a slice of $\mathrm{AdS}_5$, induced by scalars, fermions, and gauge bosons, using the well-defined and versatile Planck-brane-correlator approach, also including brane-localized fermion mass terms generically present in viable models, as well as a finite Higgs vev. A concise summary of the various contributions to the RGE is presented in Sec.~\ref{sec:summary_running}.
We also discussed the interpretation in the 5D and the dual 4D (composite Higgs) theory and commented on connections to existing results in the literature.
 
Finally, we applied our results to concrete ${\rm SU(6)} \supset {\rm SU(5)}$ GHGUTs
and presented a quantitative survey of the quality of unification for two motivated variants of fermion realizations. We found that successful grand unification at the Planck scale is possible, in the presence of moderately large brane kinetic terms $\Delta\lambda_{k,i}\approx 1.2-1.4$.

\section*{Acknowledgments}
We are grateful to Kaustubh Sadanand
Agashe, Emilian Dudas and Javier Fuentes-Martin for useful comments.

S.~W.~acknowledges support by the Cluster of Excellence ``Precision Physics, Fundamental Interactions, and Structure of Matter" (PRISMA+ EXC 2118/1) funded by the Deutsche Forschungsgemeinschaft (DFG, German Research Foundation) within the German Excellence Strategy (Project No. 390831469).
S.~W.~thanks the Max-Planck-Institut für Kernphysik for the support in the initial phase of this project.


\appendix

\section{Propagators}\label{sec:Feynman_rules}

In this appendix we derive the 5D propagators used in the main text. In the appropriate limits they agree with the results from the literature \cite{Randall:2001gb,Carena:2004zn,Falkowski:2007hz,Csaki:2010aj,Goertz:2011gk,Beneke:2012ie,Malm:2013jia}.

\subsection{Fermion Propagator}\label{sec:prop_fermion}

We start by considering the quadratic action in $\text{AdS}_{d+1}$ describing a set of fermion fields organized in a column of spinors, $\Psi = \left (\psi_1 \; \psi_2 \; \ldots  \right)^T$. Including brane masses and a non--zero $A_5$ vev, the relevant quadratic action reads
\begin{align}
    S_{d+1} &= \int d^5 x \left( \frac{1}{kz}\right)^d \overline{\Psi} \left\{ i \gamma^\mu \partial_\mu - \gamma^5 \left[\partial_z - \frac{d}{2z} - i \frac{1}{kz} g_5 \langle A_5^{\hat{a}} (z) \rangle t^{\hat{a}} \right] - \frac{c}{z} \right. \notag \\ \label{eq:ferm-5d-action}
    &\left. - \delta\left(z-\frac{1}{k}\right) \left( M_{UV} P_L + M_{UV}^\dagger P_R
    \right) - \delta\left(z-\frac{1}{T}\right) \left( M_{IR} P_L + M_{IR}^\dagger P_R
    \right)  \right\} \Psi,
\end{align}
where $t^{\hat{a}}$, $M_{UV/IR}$, and $c$ are matrices in flavor space, with $c$ being diagonal. In the absence of brane masses and $A_5$ vevs one can choose the following (Dirichlet) boundary conditions (BCs) for each field: either $P_L\psi_i(x,1/k)=0$ or $P_R\psi_i(x,1/k)=0$ and similar Dirichlet BCs for the IR brane. The BCs for the other chirality are completely determined by the equations of motion.

By defining projectors $P_{UV}$ and $P_{IR}$, who single out the LH $\psi_i$'s which have Dirichlet BCs on the UV and IR brane, respectively, they can be conveniently summarized as
\begin{align}
    &P_{UV} P_L \Psi(x,z)\rvert_{z=1/k}=0, &&P_{IR} P_L \Psi(x,z)\rvert_{z=1/T}=0, \notag \\
    (1 - &P_{UV}) P_R \Psi(x,z)\rvert_{z=1/k}=0, &(1 - &P_{IR}) P_R \Psi(x,z)\rvert_{z=1/T}=0,
    \label{eq:fermion-simple-bcs}    
\end{align}
where with a slight abuse of notation we have denoted the identity matrix in flavor space as $1$, a notation that we shall keep throughout this paper. Turning on the brane masses and the $A_5$ vevs, the boundary conditions now become:
\begin{align}\label{eq:fermion-general-bcs}
    (P_{UV} - M_{UV}) P_L \Psi(x,z)\rvert_{z=1/k} &= 0, \quad (P_{IR} + M_{IR}) \Omega\left(\frac{1}{T}\right) P_L \Psi(x,z)\rvert_{z=1/T}=0 , \notag \\
    (1 - P_{UV} + M_{UV}^\dagger) P_R \Psi(x,z)\rvert_{z=1/k} &= 0, \quad (1 - P_{IR} - M_{IR}^\dagger) \Omega\left(\frac{1}{T}\right) P_R \Psi(x,z)\rvert_{z=1/T}=0,    
\end{align}
where $\Omega(1/T)$ is the Goldstone matrix, defined as \cite{Hosotani:2006qp,Falkowski:2007hz}
\begin{equation*}\label{eq:goldstone_matrix_fermions}
    \Omega(z) \equiv \exp \left( i \, g_5 \int_{1/k}^z dz^\prime \left( \frac{1}{k\zP} \right) \langle A_5^{\hat{a}} (\zP) \rangle t^{\hat{a}}  \right), \quad \Omega(z) \Omega^\dagger(z) = \Omega^\dagger(z) \Omega(z) = 1.
\end{equation*}
Since a brane mass can only connect two fermions with $(+)$ BCs on the respective branes, it follows that $M_{UV}P_L = P_{UV} M_{UV} (1-P_{UV}) P_L$.

We now use the BCs in Eq.~\eqref{eq:fermion-general-bcs} in order to compute the (Euclidean) 5D fermion propagators. Starting from the 5D action from Eq.~\eqref{eq:ferm-5d-action} and going to Euclidean 4D momentum $p$, we get the following differential equation for the propagator $G_{1/2} (p,z,\zP)$:
\begin{equation}\label{eq:ferm-prop-diff-eq}
    \left\{ i \slashed{p} - \gamma^5 \left[\partial_z - \frac{d}{2z} - i \frac{1}{kz} g_5 \langle A_5^{\hat{a}} (z) \rangle t^{\hat{a}} \right] - \frac{c}{z} \right\}  G_{1/2} (p, z,\zP) = \left( kz \right)^d \delta(z-\zP),
\end{equation}
where he have left implicit the Dirac and flavor identity matrices on the right hand side. Also, the brane masses have been dropped as they will appear in the BCs. In order to simplify the calculations, we write the propagator as
\begin{equation}
    \label{eq:ferm-prop-simplified}
    G_{1/2} (p, z,\zP) = \left( kz \right)^{d/2} \left( k\zP \right)^{d/2} \Omega(z) \, \tilde{G}_{1/2} (p, z,\zP) \, \Omega(\zP)^\dagger.
\end{equation}
From now on, unless needed, we will omit $p$ in the argument of the propagator. Eq.~\eqref{eq:ferm-prop-diff-eq} then simplifies to:
\begin{equation}
    \label{eq:ferm-prop-diff-eq-simplified}
    \left( i \slashed{p} - \gamma^5 \partial_z - \frac{c}{z} \right) \tilde{G}_{1/2} (z,\zP) = \delta(z-\zP).
\end{equation}
Writing the propagator as
\begin{equation}\label{eq:ferm-prop-dirac-basis}
    \tilde{G}_{1/2} = P_L \left(i  \slashed{p} \, \tilde{G}_{LL} + \tilde{G}_{LR}  \right) + P_R \left(i \slashed{p} \, \tilde{G}_{RR} + \tilde{G}_{RL}  \right) ,
\end{equation}
eq.~\eqref{eq:ferm-prop-diff-eq-simplified} leads to
\begin{align}
    \label{eq:ferm-prop-chirality-diff-eqs}
    \left[ \partial_z^2 - \frac{c(c+1)}{z^2} - p^2 \right] \tilde{G}_{LL} (z,\zP) = \delta(z-\zP), \quad \tilde{G}_{RL} (z,\zP) &= \left( \partial_z + \frac{c}{z} \right) \tilde{G}_{LL} (z,\zP), \notag \\
    \left[ \partial_z^2 - \frac{c(c-1)}{z^2} - p^2 \right] \tilde{G}_{RR} (z,\zP) = \delta(z-\zP), \quad \tilde{G}_{LR} (z,\zP) &= \left(-\partial_z + \frac{c}{z} \right) \tilde{G}_{RR} (z,\zP).
\end{align}
We start by computing $\tilde{G}_{LL}$. We first write it down as
\begin{equation}
    \label{eq:GLL-separated}
    \tilde{G}_{LL} (z,\zP) = \tilde{G}_{LL}^< (z,\zP) \theta(\zP - z) + \tilde{G}_{LL}^> (z,\zP)  \theta(z-\zP),
\end{equation}
in order to separate the regions with $z>\zP$ and $z<\zP$. $\tilde{G}_{LL}$ inherits the boundary conditions from $P_L \Psi$, c.f. Eq.~\eqref{eq:fermion-general-bcs}:
\begin{align}\label{eq:fermion-prop-BCs}
    (P_{UV} - M_{UV}) \tilde{G}_{LL}^< \left(\frac{1}{k},\zP\right) &= 0, \quad (P_{IR} + M_{IR}) \Omega(1/T) \tilde{G}_{LL}^> \left(\frac{1}{T},\zP\right) = 0, \notag \\
   (1 - P_{UV} + M_{UV}^\dagger) \tilde{G}_{RL}^< \left(\frac{1}{k},\zP\right) &= 0, \quad (1 - P_{IR} - M_{IR}^\dagger) \Omega(1/T) \tilde{G}_{RL}^> \left(\frac{1}{T},\zP\right) = 0, 
\end{align}
while the junction between $\tilde{G}_{LL}^<$ and $\tilde{G}_{LL}^>$ is made through the continuity and jump conditions:
\begin{gather}
    \tilde{G}_{LL}^> (\zP,\zP) - \tilde{G}_{LL}^< (\zP,\zP) = 0 \notag \\
    \label{eq:GLL-cont-cdt}
    \left[ \left( \partial_z + \frac{c}{z} \right) \tilde{G}_{LL}^> (z,\zP) - \left( \partial_z + \frac{c}{z} \right) \tilde{G}_{LL}^< (z,\zP)  \right]_{z=\zP} = 1,
\end{gather}
where in the second line we have introduced the $c/z$ terms for later convenience, even though they cancel out by virtue of the continuity condition from the first line. 

To simplify the calculations, we derive a property of $G_{LL}$ using $G_{LL}(z,\zP) \propto \expval{\Psi_L(z)\overline{\Psi_L}(\zP)}$. It follows immediately that $\tilde{G}_{LL}^>$ and $\tilde{G}_{LL}^<$ can be related in a simple way:
\begin{equation}
    \label{eq:prop-symmetry-relation}
    \tilde{G}_{LL} (z,\zP) = \tilde{G}_{LL}^\dagger (\zP,z) \: \Rightarrow \: \tilde{G}_{LL}^> (z,\zP) = \tilde{G}_{LL}^{<,\dagger} (\zP,z).
\end{equation}
Later on we shall use this relation\footnote{As we have explicitly checked, one can reach the same results without resorting to this symmetry relation, albeit with much more effort.} to derive a simple but general expression for fermion propagators in the presence of brane masses and $A_5$ vevs.

We now define two functions,
\begin{align}\label{eq:warped-trig-fcts-tilde}
    \tilde{C}_c^{k} (z) &= \frac{p}{k} \sqrt{kz} \left[ K_{c-1/2} (p/k) I_{c+1/2} (p z) + I_{c-1/2} (p/k) K_{c+1/2} (p z) \right], \notag \\
    \tilde{S}_c^{k} (z) &= \frac{p}{k} \sqrt{kz} \left[ K_{c+1/2} (p/k) I_{c+1/2} (p z) - I_{c+1/2} (p/k) K_{c+1/2} (p z) \right],
\end{align}
which both solve the differential equation for $G_{LL} (z,\zP)$ with $z \neq \zP$, see Eq.~\eqref{eq:ferm-prop-chirality-diff-eqs}. They are similar to the warped sine and cosine defined in \cite{Falkowski:2006vi}, and satisfy several properties reminiscent of the hyperbolic sine and cosine, to which they reduce for $c=0$:
\begin{gather}
    \tilde{C}_0^{k} (z) = \cosh \left[ p (z-1/k) \right], \quad \tilde{S}_0^{k} (z) = \sinh \left[ p (z-1/k) \right], \notag \\
    \tilde{C}_c^{k} (1/k) = 1, \quad \tilde{S}_c^{k} (1/k) = 0, \quad \tilde{C}_c^{k} (z) \tilde{C}_{-c}^{k} (z) - \tilde{S}_c^{k} (z) \tilde{S}_{-c}^{k} (z) = 1, \notag \\ \label{eq:warped-cosh-sinh-properties}
    \left( \partial_z + \frac{c}{z} \right) \tilde{C}_c^{k} (z) = p \, \tilde{S}_{-c}^{k} (z), \quad \left( \partial_z + \frac{c}{z} \right) \tilde{S}_c^{k} (z) = p \, \tilde{C}_{-c}^{k} (z).
\end{gather}
Similarly, we define $\tilde{C}_c^{T}$ and $\tilde{S}_c^{T}$ with $k\to T$. With the help of these functions, we define
\begin{align}\label{eq:ftilde-definitions}
    \tilde{f}_L^{k} (z) &= \tilde{S}_c^{k} (z) (P_{UV}- M_{UV}^\dagger) + \tilde{C}_c^{k} (z) (1 - P_{UV} + M_{UV}), \notag \\ 
    \tilde{f}_L^{T} (z) &= \tilde{S}_c^{T} (z) (P_{IR} + M_{IR}^\dagger) + \tilde{C}_c^{T} (z) (1 - P_{IR} - M_{IR}), \notag \\
   \frac{1}{p} \left( \partial_z + \frac{c}{z} \right)  \tilde{f}_L^{k} (z) \equiv \tilde{f}_R^{k} (z) &= \tilde{C}_{-c}^{k} (z) (P_{UV}- M_{UV}^\dagger) + \tilde{S}_{-c}^{k} (z) (1 - P_{UV} + M_{UV}), \notag \\
   \frac{1}{p} \left( \partial_z + \frac{c}{z} \right)  \tilde{f}_L^{T} (z) \equiv \tilde{f}_R^{T} (z) &= \tilde{C}_{-c}^{T} (z) (P_{IR} + M_{IR}^\dagger) + \tilde{S}_{-c}^{T} (z) (1 - P_{IR} - M_{IR}).
\end{align}
Using the brane mass matrix properties from Eq.~\eqref{eq:fermion-prop-BCs} and the warped hyperbolic functions' properties from Eq.~\eqref{eq:warped-cosh-sinh-properties}, it is straightforward to show that $\tilde{f}_L$ satisfy almost the same BCs as $\tilde{G}_{LL}$.

One can now write an ansatz for $\tilde{G}_{LL}$ which satisfies both the BCs from Eq.~\eqref{eq:fermion-prop-BCs} and the symmetry property from Eq.~\eqref{eq:prop-symmetry-relation}:
\begin{align}
    \tilde{G}_{LL}^< (z,\zP) &= \frac{1}{p} \tilde{f}_L^k (z) \, \mathcal{M}_{\frac{1}{2}} \, \tilde{f}_L^{T,\dagger} (\zP) \, \Omega(1/T), \notag \\ \label{eq:propagator_ansatz}
    \tilde{G}_{LL}^{>} (z,\zP) &= \frac{1}{p} \Omega^\dagger(1/T) \, \tilde{f}_L^{T} (z) \, \mathcal{M}_{\frac{1}{2}}^\dagger \, \tilde{f}_L^{k,\dagger} (\zP),
\end{align}
where $\mathcal{M}_{\frac{1}{2}}$ is a matrix in flavor space. Assuming $\mathcal{M}_{\frac{1}{2}}$ does not depend on $z,\zP$, $\mathcal{M}_{\frac{1}{2}}$ can be computed by introducing the propagator ansatz from above into the continuity and jump conditions in Eq.~\eqref{eq:GLL-cont-cdt}:
\begin{align}
    \Omega^\dagger(1/T) \, \tilde{f}_L^{1/T} (\zP) \, \mathcal{M}_{\frac{1}{2}}^\dagger \, \tilde{f}_L^{k,\dagger} (\zP) - \tilde{f}_L^{k} (\zP) \, \mathcal{M}_{\frac{1}{2}} \, \tilde{f}_L^{T,\dagger} (\zP) \, \Omega(1/T) &= 0, \notag \\
    \Omega^\dagger(1/T) \, \tilde{f}_R^{1/T} (\zP) \, \mathcal{M}_{\frac{1}{2}}^\dagger \, \tilde{f}_L^{k,\dagger} (\zP) - \tilde{f}_R^{k} (\zP) \, \mathcal{M}_{\frac{1}{2}} \, \tilde{f}_L^{1/T,\dagger} (\zP) \, \Omega(1/T) &= 1,
\end{align}
where we have used the definitions of $\tilde{f}_R$ from Eq.~\eqref{eq:ftilde-definitions} and the fact that $\Omega(1/T)$ commutes with $c$.\footnote{The fact that $\Omega(z)$ and $\Omega^\dagger (z)$ commute with the diagonal matrix $c$ follows from the fact that that the $A_5$ vev only mixes fermion fields from the same bulk multiplet, in other words only fermions having the same $c$ parameter.} From here, the expression for $\mathcal{M}_{\frac{1}{2}}$ follows:
\begin{equation}
    \mathcal{M}_{\frac{1}{2}}^{-1} = f_R^{T,\dagger}(\zP) \Omega(1/T) \tilde{f}_L^{k}(\zP) - \tilde{f}_L^{T,\dagger}(\zP) \Omega(1/T) \tilde{f}_R^{k}(\zP) ,
\end{equation}
where we used that $\tilde{f}_R^\dagger (\zP) \tilde{f}_L (\zP) = \tilde{f}_L^\dagger (\zP) \tilde{f}_R (\zP)$. An explicit computation verifies that $\partial_z \mathcal{M}_{\frac{1}{2}}^{-1}=0$ and it will be convenient to evaluate $\mathcal{M}_{\frac{1}{2}}$ at $\zP=1/T$
\begin{equation}
    \label{eq:m-final-expression}
    \mathcal{M}_{\frac{1}{2}}^{-1} = (P_{IR} + M_{IR})  \Omega(1/T) \tilde{f}_L^{k} (1/T) - (1 - P_{IR} - M_{IR}^\dagger ) \Omega(1/T) \tilde{f}_R^k (1/T).
\end{equation}
A useful consistency check of the above result is provided by observing that the fermionic spectral function is, up to an unphysical sign, given by:
\begin{equation*}
    \rho_{1/2}(-p^2) = \det \left[ (P_{IR} + M_{IR})  \Omega(1/T) \tilde{f}_L^k (1/T) - (1 - P_{IR} - M_{IR}^\dagger ) \Omega(1/T) \tilde{f}_R^k (1/T) \right],
\end{equation*}
which guarantees that the Euclidean propagator enjoys the correct property of having poles along the imaginary axis at $p^2 = - m_n^2$, with $m_n$ denoting the fermion KK spectrum.

One can repeat the same exercise to compute $\tilde{G}_{RR}$, while $\tilde{G}_{LR}$ and $\tilde{G}_{RL}$ follow immediately from Eq.~\eqref{eq:ferm-prop-chirality-diff-eqs}. For the reader's convenience, we summarize below all the results for the propagators:
\begin{align}
   p \, \tilde{G}_{LL}^< (z,\zP) &= \tilde{f}_L^k (z) \, \mathcal{M} \, \tilde{f}_L^{T,\dagger} (\zP) \, \Omega(1/T), \quad \tilde{G}_{LL}^{>} (z,\zP) = \tilde{G}_{LL}^{<,\dagger} (\zP,z), \notag \\
   \tilde{G}_{LR}^< (z,\zP) &=  \tilde{f}_L^k (z) \, \mathcal{M} \, \tilde{f}_R^{T,\dagger} (\zP) \, \Omega(1/T), \quad \tilde{G}_{LR}^{>} (z,\zP) = \tilde{G}_{RL}^{<,\dagger} (\zP,z), \notag \\
   -p \, \tilde{G}_{RR}^< (z,\zP) &= \tilde{f}_R^k (z) \, \mathcal{M} \, \tilde{f}_R^{T,\dagger} (\zP) \, \Omega(1/T), \quad \tilde{G}_{RR}^{>} (z,\zP) = \tilde{G}_{RR}^{<,\dagger} (\zP,z), \notag \\
   \tilde{G}_{RL}^< (z,\zP) &=  \tilde{f}_R^k (z) \, \mathcal{M} \, \tilde{f}_L^{T,\dagger} (\zP) \, \Omega(1/T), \quad \tilde{G}_{RL}^{>} (z,\zP) = \tilde{G}_{LR}^{<,\dagger} (\zP,z),
\end{align}
with $\tilde{f}_{L/R}$ and $\mathcal{M}$ given in Eqs.~\eqref{eq:ftilde-definitions} and \eqref{eq:m-final-expression}, respectively. This completes the calculation of the fermion propagator.

\subsection{Scalar Propagator}\label{sec:prop_scalar}

One can use a similar derivation as in the previous section for scalar propagators. We start by considering the quadratic action describing a set of complex scalar fields organized in a column of scalars, $\Phi = \left (\phi_1 \; \phi_2 \; \ldots  \right)^T$ in $\text{AdS}_{d+1}$. As we saw in Section~\ref{sec:derivation}, the dynamics on the IR brane are irrelevant for the differential running, so we neglect the gauge-Higgs vev and IR brane masses and only include UV brane masses in the following. With this, the relevant quadratic action reads
\begin{align}\label{eq:scalar-action}
    S_{d+1} &= \int d^d x \int d z \left( \frac{1}{kz}\right)^{d-1} \Phi^{\dagger} \left\{ - \partial^2 + \left(kz\right)^{d-1}\partial_z\left(\frac{1}{kz}\right)^{d-1}\partial_z \right. \notag \\
    &\left. \hspace{5cm} - \left(\frac{1}{kz}\right)^{2}m^2 + \delta(z-1/k) M_{UV}^2k \right\} \Phi.
\end{align}
In the absence of brane masses the fields satisfy either Dirichlet BCs on the UV brane $\phi_{i}(x,z)\rvert_{z=1/k}=0$ or von Neumnann BCs $\partial_z\phi_{i}(x,z)\rvert_{z=1/k}=0$ and similar conditions for the IR brane. By defining projectors $P_{UV}$ and $P_{IR}$, who single out the $\phi_i$'s which have Dirichlet BCs on the UV and IR brane, respectively, they can be conveniently summarized as
\begin{align}
    &P_{UV} \Phi(x,z)\rvert_{z=1/k}=0, &&P_{IR} \Phi(x,z)\rvert_{z=1/T}=0, \notag \\
    (1 - &P_{UV}) \partial_z \Phi(x,z)\rvert_{z=1/k}=0, &(1 - &P_{IR}) \partial_z \Phi(x,z)\rvert_{z=1/T}=0,
    \label{eq:scalar-simple-bcs}    
\end{align}
Turning on the brane masses, the boundary conditions now become:
\begin{align}
    &P_{UV} \Phi(x,z)\rvert_{z=1/k}=0, &&P_{IR} \Phi(x,z)\rvert_{z=1/T}=0, \notag \\
    (1 - &P_{UV}) \left(\partial_z-M_{UV}^2k\right) \Phi(x,z)\rvert_{z=1/k}=0, &(1 - &P_{IR}) \partial_z \Phi(x,z)\rvert_{z=1/T}=0.
    \label{eq:scalar-general-bcs}    
\end{align}

The quadratic action \eqref{eq:scalar-action} and the BCs \eqref{eq:scalar-general-bcs} can be used to determine the (Euclidean) 5D scalar field propagator. Starting from the 5D action from \eqref{eq:scalar-action} and going to Euclidean 4D momentum $p \to i p$, we get the following differential equation for the propagator $G_{0} (p,z,\zP)$:
\begin{equation}\label{eq:scalar-prop-diff-eq}
    \left\{ -p^2 + \left(kz\right)^{d-1}\partial_z\left(\frac{1}{kz}\right)^{d-1}\partial_z - \left(\frac{1}{kz}\right)^{2}m^2 \right\} G_{0} (p, z,\zP) = \left( kz \right)^{d-1} \delta(z-\zP)
\end{equation}
and the BCs
\begin{align}
    &P_{UV} G_{0}^{<}(p, z,\zP)(x,z)\rvert_{z=1/k}=0, &&P_{IR} G_{0}^{>} (p, z,\zP)\rvert_{z=1/T}=0, \notag \\
    (1 - &P_{UV}) \left(\partial_z-M_{UV}^2k\right) G_{0}^{<} (p, z,\zP)\rvert_{z=1/k}=0, &(1 - &P_{IR}) \partial_z G_{0}^{>} (p, z,\zP)\rvert_{z=1/T}=0.
\end{align}
Here we have already written the propagators as
\begin{equation}
    G_{0}(p,z,\zP) = G_{0}^{<}(p,z,\zP) \theta(\zP-z) + G_{0}^{>}(p,z,\zP) \theta(z-\zP).
\end{equation}
Each of them satisfies the differential equation \eqref{eq:scalar-prop-diff-eq} with the right hand side set to zero and a continuity and jump condition for the propagator at $z=\zP$ following from that equation which we will specify later. Moreover, since $G_{0}(p,z,\zP) \propto \expval{\Phi(z)\Phi^{\dagger}(\zP)}$, we can relate these two via
\begin{equation}
    G_{0}(p,z,\zP) = G_{0}^{\dagger}(p,\zP,z) \: \Rightarrow \: G_{0}^{>}(p,z,\zP) = G_{0}^{<,\dagger}(p,\zP,z)\,.
\end{equation}
To simplify the differential equation we first write
\begin{equation}
    G_{0}(p,z,\zP) = \left(kz\right)^{\frac{d-1}{2}} \left(k\zP\right)^{\frac{d-1}{2}} \tilde{G}_{0}(z,\zP), 
\end{equation}
where we will from now on omit $p$ in the argument of the propagator unless needed. Then, \eqref{eq:scalar-prop-diff-eq} simplifies to
\begin{equation}\label{eq:scalar-prop-diff-eq_simplified}
    \left\{ \partial_z^2 - \frac{\alpha(\alpha+1)}{z^2} -p^2 \right\} \tilde{G}_{0} (z,\zP) = \delta(z-\zP),
\end{equation}
where we defined $\alpha=\frac{1}{2}(\sqrt{d^2+4m^2/k^2}-1)$. The functions
\begin{align}\label{eq:scalar-warped-trigs}
\tilde{C}_{\alpha}^{k} (z) &= \frac{p}{k} \sqrt{kz} \left[ A_{\alpha-\frac{1}{2}}(p/k) I_{\alpha+\frac{1}{2}} (p z) + B_{\alpha-\frac{1}{2}}(p/k) K_{\alpha+\frac{1}{2}} (p z) \right], \notag \\
\tilde{S}_{\alpha}^{k} (z) &= \frac{p}{k} \sqrt{kz} \left[ K_{\alpha+\frac{1}{2}} (p/k) I_{\alpha+\frac{1}{2}} (p z) - I_{\alpha+\frac{1}{2}} (p/k) K_{\alpha+\frac{1}{2}} (p z) \right], \notag \\
\tilde{D}_{\alpha}^{k} (z) &= \frac{p}{k} \sqrt{kz} \left[ A_{\alpha-\frac{1}{2}}(p/k) B_{\alpha-\frac{1}{2}} (p z) - B_{\alpha-\frac{1}{2}}(p/k) A_{\alpha-\frac{1}{2}} (p z) \right], \notag \\
\tilde{T}_{\alpha}^{k} (z) &= \frac{p}{k} \sqrt{kz} \left[ K_{\alpha+\frac{1}{2}} (p/k) B_{\alpha-\frac{1}{2}} (p z) + I_{\alpha+\frac{1}{2}} (p/k) A_{\alpha-\frac{1}{2}} (p z) \right],
\end{align}
with
\begin{align}
    A_{\alpha-\frac{1}{2}}(p/k) &= K_{\alpha+\frac{3}{2}}(p/k) - \frac{(\alpha+\frac{1}{2})+(\beta+\frac{1}{2})}{p/k} K_{\alpha+\frac{1}{2}}(p/k), \notag \\
    B_{\alpha-\frac{1}{2}}(pR) &= I_{\alpha+\frac{3}{2}}(pR) + \frac{(\alpha+\frac{1}{2})+(\beta+\frac{1}{2})}{p/k} I_{\alpha+\frac{1}{2}}(p/k),
\end{align}
solve the differential equation for $z\neq\zP$. We have defined $\beta = \frac{d-1}{2}$ and one has $A_{\alpha-\frac{1}{2}}(p/k) \to K_{\alpha-\frac{1}{2}}(p/k)$ and $B_{\alpha-\frac{1}{2}}(p/k) \to I_{\alpha-\frac{1}{2}}(p/k)$ for $m^2\to 0$ or equivalently $\alpha \to \beta$. We normalized these functions such that they satisfy
\begin{gather}\label{eq:scalar-warped-trigs-properties}
    \tilde{C}_{\alpha}^{k} (1/k) = 1, \quad \tilde{S}_{\alpha}^{k} (1/k) = 0, \quad \tilde{D}_{\alpha}^{k} (1/k) = 0, \quad \tilde{T}_{\alpha}^{k} (1/k) = 1, \notag \\ 
    \left( \partial_z + \frac{\beta}{z} \right) \tilde{C}_{\alpha}^{k} (z) = p \, \tilde{D}_{\alpha}^{k} (z), \quad \left( \partial_z + \frac{\beta}{z} \right) \tilde{S}_{\alpha}^{k} (z) = p \, \tilde{T}_{\alpha}^{k} (z).
\end{gather}
We can define similarly $\tilde{C}_{\alpha}^{T},\tilde{S}_{\alpha}^{T},\tilde{D}_{\alpha}^{T},\tilde{T}_{\alpha}^{T}$ by replacing $k \to T$. With these we define now
\begin{align}\label{eq:scalar-ftilde-definitions}
    \tilde{h}_L^{k} (z) &= \tilde{S}_{\alpha}^{k} (z) P_{UV} + \left[\tilde{C}_{\alpha}^{k} (z) + M_{UV}^2 \frac{\tilde{S}_{\alpha}^{k} (z)}{p/k} \right](1 - P_{UV}), \notag \\ 
    \tilde{h}_L^{T} (z) &= \tilde{S}_{\alpha}^{T} (z) P_{IR} + \tilde{C}_{\alpha}^{T} (z) (1 - P_{IR}), \notag \\
    \tilde{h}_R^{k} (z) &= \tilde{T}_{\alpha}^{k} (z) P_{UV} + \left[\tilde{D}_{\alpha}^{k} (z) + M_{UV}^2 \frac{\tilde{T}_{\alpha}^{k} (z)}{p/k} \right] (1 - P_{UV}), \notag \\ 
    \tilde{h}_R^{T} (z) &= \tilde{T}_{\alpha}^{T} (z) P_{IR} + \tilde{D}_{\alpha}^{T} (z) (1 - P_{IR}),
\end{align}
which also satisfy the BCs. Here the index $L/R$ is introduced to show the similarities with the fermion case. These functions satisfy
\begin{align}
    \frac{1}{p} \left( \partial_z + \frac{\beta}{z} \right)  \tilde{h}_L (z) &= \tilde{h}_R (z)
\end{align}
We can write down explicitly the solution for $\tilde{G}_{0}$ as
\begin{align}
    \tilde{G}_{0}^{<}(z,\zP) &= \frac{1}{p} \tilde{h}_L^{k} (z) \mathcal{M}_{0} \tilde{h}_L^{T,\dagger} (\zP), \notag \\
    \tilde{G}_{0}^{>}(z,\zP) &= \frac{1}{p} \tilde{h}_L^{T} (z) \mathcal{M}_{0}^{\dagger} \tilde{h}_L^{k,\dagger} (\zP),
\end{align}
where $\mathcal{M}_0$ is a matrix in flavor space introduced such that $\tilde{G_{0}}$ fulfills the continuity and jump conditions
\begin{gather}\label{eq:scalar-cont-cdts}
    \tilde{G}_{0}^> (\zP,\zP) - \tilde{G}_{0}^< (\zP,\zP) = 0 \notag \\
    \left[ \left( \partial_z + \frac{\beta}{z} \right) \tilde{G}_{0}^> (z,\zP) - \left( \partial_z + \frac{\beta}{z} \right) \tilde{G}_{0}^< (z,\zP)  \right]_{z=\zP} = 1.
\end{gather}
On can solve these equations for $\mathcal{M}_0$ under the assumption that $\mathcal{M}_0$ does not depend on $z,\zP$ and using $\tilde{h}_L^{\dagger}(\zP)\tilde{h}_R(\zP)=\tilde{h}_R^{\dagger}(\zP)\tilde{h}_L(\zP)$, giving
\begin{equation}
   \mathcal{M}_{0}^{-1} = \tilde{h}_R^{T,\dagger}(\zP)\tilde{h}_L^{k}(\zP) - \tilde{h}_L^{T,\dagger}(\zP)\tilde{h}_R^{k}(\zP) .
\end{equation}
An explicit computation verifies $\partial_{\zP}\mathcal{M}_{0}^{-1}=0$ and it will be convenient to evaluate it at $\zP=1/T$ giving
\begin{equation}
    \mathcal{M}_{0}^{-1} = P_{IR}\tilde{h}_L^{k}(1/T) - (1-P_{IR})\tilde{h}_R^{k}(1/T) .
\end{equation}
This completes the calculation of the scalar propagator.

\subsection{Gauge boson Propagator}\label{sec:prop_gauge_boson}

One can use once more the same approach in order to compute the gauge boson propagator. We start by considering the quadratic action in $\text{AdS}_{d+1}$ describing a set of gauge boson fields organized in a column of gauge fields, $A_{\mu} = \left (A_{\mu}^1 \; A_{\mu}^2 \; \ldots  \right)^T$. Including UV-brane masses and a non--zero $A_5$ vev, the relevant quadratic action reads
\begin{align}
    S_{d+1} &= \int d^5 x \left(\frac{1}{kz}\right)^{d-3} A_{\mu} \frac{1}{2}\left\{ -\eta^{\mu\nu}\partial^2 + \left(1-\frac{1}{\xi}\right)\partial^{\mu}\partial^{\nu} +\eta^{\mu\nu} \left(kz\right)^{d-3} \partial_z \left(\frac{1}{kz}\right)^{d-3} \partial_z \right. \notag \\ \label{eq:gauge-5d-action}
    &\left. \hspace{5cm}+ \eta^{\mu\nu}\mathcal{M}^2 + \delta(z-1/k) \eta^{\mu\nu} \left(M_{UV}^2k\right)  \right\} A_{\nu},
\end{align}
with $\mathcal{M}_{ab}^2=2g_5^2 f^{Ea\hat{c}}f^{Eb\hat{d}}\langle A_{5}^{\hat{c}}(z)\rangle \langle A_{5}^{\hat{d}}(z)\rangle$, while $M_{UV}^2$ is a matrix in gauge space with dimensionless entries. If sourced by a UV brane scalar getting a vev $v_{GUT} \sim M_{GUT}$, the typical values for these entries are $ (g_5^2k) k^{-2} M_{GUT}^2 = g_*^2 M_{GUT}^2/M_{Pl}^2$, therefore rather suppressed.

In the absence of brane masses and Higgs vev the fields satisfy either Dirichlet BCs on the UV brane $A_{\mu}^i(x,z)\rvert_{z=1/k}=0$ or von Neumnann BCs $\partial_z A_{\mu}^i(x,z)\rvert_{z=1/k}=0$ and similar conditions for the IR brane. By defining projectors $P_{UV}$ and $P_{IR}$, who single out the $A_{\mu}^i$'s which have Dirichlet BCs on the UV and IR brane, respectively, they can be conveniently summarized as
\begin{align}
    &P_{UV} A_{\mu}(x,z)\rvert_{z=1/k}=0, &&P_{IR} A_{\mu}(x,z)\rvert_{z=1/T}=0, \notag \\
    (1 - &P_{UV}) \partial_z A_{\mu}(x,z)\rvert_{z=1/k}=0, &(1 - &P_{IR}) \partial_z A_{\mu}(x,z)\rvert_{z=1/T}=0\,.
    \label{eq:gauge-simple-bcs}    
\end{align}
Turning on the brane masses and the $A_5$ vevs, the boundary conditions become
\begin{align}
    &P_{UV} A_{\mu}(x,z)\rvert_{z=1/k}=0, &&P_{IR} \Omega(1/T) A_{\mu}(x,z)\rvert_{z=1/T}=0, \notag \\
    (1 - &P_{UV}) \left(\partial_z-M_{UV}^2k\right) A_{\mu}(x,z)\rvert_{z=1/k}=0, &(1 - &P_{IR}) \Omega(1/T) \partial_z A_{\mu}(x,z)\rvert_{z=1/T}=0.
    \label{eq:gauge-general-bcs}    
\end{align}
Note that a brane mass can only connect fields which are non--zero at the respective boundary giving $M_{UV}^2=(1-P_{UV})M_{UV}^2(1-P_{UV})$.

We now define the Goldstone matrix leading to mixing between the gauge bosons:
\begin{equation}
    \label{eq:goldstone-matrix-gauge}
    \Omega_{ab}(z) = 2 \Tr{ \Omega_f(z)  T^b \Omega_f^\dagger(z) T^a }, \quad
\end{equation}
where 
\begin{equation*}
\Omega_f (z) = \exp \left( i \, g_5 \int_{1/k}^z dz^\prime \left( \frac{1}{k\zP} \right) \langle A_5^{\hat{c}} (\zP) \rangle T^{\hat{c}}  \right)    
\end{equation*}
is the Goldstone matrix given in Eq.~\eqref{eq:goldstone_matrix_fermions} pertaining to the fundamental representation and the $T$'s indicate the SU(N) generators in the fundamental representation. 

Starting from the 5D action from \eqref{eq:gauge-5d-action} and going to Euclidean 4D momentum $p \to i p$, we get the following differential equation for the propagator $G_{1,\nu\lambda} (p,z,\zP)$:
\begin{align}\label{eq:gauge-prop-diff-eq}
    &\left\{ -\delta^{\mu\nu}p^2 + \left(1-\frac{1}{\xi}\right)p^{\mu}p^{\nu} +\delta^{\mu\nu} \left(kz\right)^{d-3} \partial_z \left(\frac{1}{kz}\right)^{d-3} \partial_z + \delta^{\mu\nu}\mathcal{M}^2\right\} G_{1,\nu\lambda} (p, z,\zP) \nonumber \\
    &\hspace{10cm}= \left( kz \right)^{d-3} \delta(z-\zP) \delta_{\lambda}^{\mu}.
\end{align}
The tensor structure as well as the gauge dependence can be simplified by the following ansatz \cite{Randall:2001gb}
\begin{equation}\label{eq:gauge_prop_tensor_reduction}
   G_{1,\nu\lambda}(p,z,\zP) = G_1(p,z,\zP) \left(\delta_{\nu\lambda}-\frac{p_{\nu}p_{\lambda}}{p^2}\right) + G_1\left(\frac{p}{\sqrt{\xi}},z,\zP\right) \left(\frac{p_{\nu}p_{\lambda}}{p^2}\right) .
\end{equation}
Furthermore, we write the gauge propagator as (dropping $p$ from the argument of the propagator in order to simplify notation):
\begin{equation}
    \label{eq:gauge-prop-simplified}
    G_1 (z,\zP) = \left( kz \right)^{(d-3)/2} \left( k\zP \right)^{(d-3)/2} \Omega(z) \, \tilde{G}_{1} ( z,\zP) \, \Omega(\zP)^\dagger,
\end{equation}
such that $\tilde{G}_1$ satisfies
\begin{equation}
    \label{eq:gauge-prop-simplified-diff-eqs}
     \left[ \partial_z^2 - \frac{\alpha(\alpha+1)}{z^2}  - p^2 \right] \tilde{G}_1 (z,\zP) = -\delta(z-\zP),
\end{equation}
where we have defined $\alpha = \frac{d-3}{2}$. Up to a minus sign, $ \tilde{G}_1 (z,\zP)$ satisfies the same differential equation as $\tilde{G}_{LL} (z,\zP)$ (see Eq.~\eqref{eq:ferm-prop-chirality-diff-eqs}) with $c=\alpha$.  With this, the propagator BCs become:
\begin{gather}
P_{UV} \tilde{G}_1^< (1/k,\zP) = 0, \quad (1-P_{UV})\left( \partial_z +\frac{\alpha}{z} -M_{UV}^2k \right) \tilde{G}_1^< (z,\zP) \Big|_{z=1/k} = 0, \notag \\ \label{eq:BCs-gauge-prop}
P_{IR} \Omega(1/T) \tilde{G}_1^> (1/T,\zP) = 0, \quad (1-P_{IR}) \Omega(1/T) \left( \partial_z +\frac{\alpha}{z}\right) \tilde{G}_1^> (z,\zP) \Big|_{z=1/T} = 0.
\end{gather}

The rest of the derivation proceeds as in the fermionic case. We therefore quote just the final result:
\begin{equation}
    \label{eq:gauge-prop-final-result}
    \tilde{G}_1^< (z,\zP) = -\frac{1}{p} \tilde{g}_L^{k} (z) \, \mathcal{M}_{1} \, \tilde{g}_L^{T} (\zP) \, \Omega(1/T), \quad \tilde{G}_1^> (z,\zP) = \tilde{G}_1^{<,\dagger} (\zP,z),
\end{equation}
with ($\rho_1$ is the gauge boson spectral function):
\begin{equation}
    \mathcal{M}_{1}^{-1} = P_{IR} \Omega(1/T) \tilde{g}_L^k (1/T) - (1-P_{IR}) \Omega(1/T) \tilde{g}_R^k (1/T) \quad \Rightarrow \quad \rho_1 (-p^2) = \det \mathcal{M}_{1}^{-1}.
\end{equation}
In the equations above, we have used the following shorthand notations:
\begin{align}
    \tilde{g}_L^k (z) &= \tilde{S}_{\alpha}^{k} (z) P_{UV} + \left[\tilde{C}_{\alpha}^k (z) + M_{UV}^2 \frac{\tilde{S}_{\alpha}^{k} (z) }{p/k} \right] (1 - P_{UV}), \notag \\ 
    \tilde{g}_L^{T} (z) &= \tilde{S}_{\alpha}^{T} (z) P_{IR} + \tilde{C}_{\alpha}^{T} (z) (1 - P_{IR}), \notag \\
    \tilde{g}_R^k (z) &= \tilde{C}_{-\alpha}^k (z) P_{UV} + \left[\tilde{S}_{-\alpha}^k (z) + M_{UV}^2 \frac{\tilde{C}_{-\alpha}^k (z)}{p/k} \right] (1 - P_{UV}), \notag \\
    \tilde{g}_R^{T} (z) &= \tilde{C}_{-\alpha}^{T} (z) P_{IR} + \tilde{S}_{-\alpha}^{T} (z) (1 - P_{IR}) .
    \label{eq:gtilde-definitions}
\end{align}
This completes the calculation of the gauge boson propagator.

\subsection{Ghost Propagator}\label{sec:prop_ghost}

We saw in Appendix~\ref{sec:prop_gauge_boson} that the effect of the Gauge-Higgs vev modifies the IR BC after a suitable gauge transformation. Moreover, we saw in Section~\ref{sec:derivation} that the dynamics on the IR brane are irrelevant for the running of the Planck-brane correlator. To simplify the ghost field propagator we drop therefore, the Gauge-Higgs vev and IR brane masses in this section. Since none of the models we consider in this work has UV brane masses for gauge bosons we also drop them in the following. With these assumptions the gauge fixing functional reduces to known expressions~\cite{Randall:2001gb} and by varying it, we obtain the quadratic part of the ghost action
\begin{align}
    S_{d+1} &= \int d^5 x \left(\frac{1}{kz}\right)^{d-3} \Bar{c}\left\{ \partial^2 +\xi \left(kz\right)^{d-3} \partial_z \left(\frac{1}{kz}\right)^{d-3} \partial_z\right\} c \label{eq:ghost-5d-action} \,.
\end{align}
The boundary conditions are the same as the ones for the corresponding gauge fields $A_{\mu}$ and we can write
\begin{align}
    &P_{UV} c(x,z)\rvert_{z=1/k}=0, &&P_{IR} c(x,z)\rvert_{z=1/T}=0, \notag \\
    (1 - &P_{UV}) \partial_z c(x,z)\rvert_{z=1/k}=0, &(1 - &P_{IR}) \partial_z c(x,z)\rvert_{z=1/T}=0,
    \label{eq:ghost-simple-bcs}    
\end{align}
with the same $P_{UV}$ and $P_{IR}$ as in Section~\ref{sec:prop_gauge_boson}.

Starting from the 5D action from \eqref{eq:ghost-5d-action} and going to Euclidean 4D momentum $p \to i p$, we get the following differential equation for the propagator $G_{g} (p,z,\zP)$:
\begin{equation}\label{eq:ghost-prop-diff-eq}
    \left\{ -p^2 +\xi \left(kz\right)^{d-3} \partial_z \left(\frac{1}{kz}\right)^{d-3} \partial_z \right\} G_{g} (p, z,\zP) = \left( kz \right)^{d-3} \delta(z-\zP).
\end{equation}
Comparing this to Eq.~\eqref{eq:gauge-prop-diff-eq} for gauge fields and because the ghost fields have the same BCs as the gauge fields we can immediately write down the solution of the ghost propagator as~\cite{Randall:2001gb}
\begin{equation}\label{eq:ghost_prop_solution}
   G_{g}(p,z,\zP) = \frac{1}{\xi}G_1\left(\frac{p}{\sqrt{\xi}},z,\zP\right) ,
\end{equation}
with the expression for $G_1(p/\sqrt{\xi},z,\zP)$ given in Appendix~\ref{sec:prop_gauge_boson}.

\subsection{\texorpdfstring{$A_5$}{A5} Propagator}\label{sec:prop_A5}

For $A_5$, we neglect again the gauge-Higgs vev and brane masses on the IR brane as they are not relevant to the running. We also do not use any UV brane mass as none of the models considered in this work contains one. With this the quadratic part of the action for $A_5$ reads
\begin{align}
    S_{d+1} &= \int d^5 x \left(\frac{1}{kz}\right)^{d-3} A_{5} \frac{1}{2}\left\{ -\partial^2 +\xi \partial_z\left(kz\right)^{d-3} \partial_z \left(\frac{1}{kz}\right)^{d-3} \right\} A_{5} \label{eq:scalar-gauge-5d-action}\,.
\end{align}
The BCs are opposite to the BCs for the corresponding gauge fields $A_{\mu}$ and thus can be written as
\begin{align}
    (1-&P_{UV}) A_{5}(x,z)\rvert_{z=1/k}=0, &(1-&P_{IR}) A_{5}(x,z)\rvert_{z=1/T}=0, \notag \\
    &P_{UV} \partial_z \left(\left(\frac{1}{kz}\right)^{d-3}A_{5}(x,z)\right)\bigg\rvert_{z=1/k}=0, &&P_{IR} \partial_z \left(\left(\frac{1}{kz}\right)^{d-3} A_{5}(x,z)\right)\bigg\rvert_{z=1/T}=0,
    \label{eq:scalar-gauge-simple-bcs}    
\end{align}
with the same $P_{UV}$ and $P_{IR}$ as in in Section~\ref{sec:prop_gauge_boson}.

Starting from the 5D action from \eqref{eq:scalar-gauge-5d-action} and going to Euclidean 4D momentum $p \to i p$, we get the following differential equation for the propagator $G_{5} (p,z,\zP)$:
\begin{equation}\label{eq:A5-prop-diff-eq}
    \left\{ -p^2 +\xi \partial_z\left(kz\right)^{d-3} \partial_z \left(\frac{1}{kz}\right)^{d-3} \right\} G_{5} (p, z,\zP) = \left( kz \right)^{d-3} \delta(z-\zP).
\end{equation}
We write the $A_5$ propagator as (dropping $p$ from the argument of the propagator in order to simplify notation):
\begin{equation}
    \label{eq:A5-prop-simplified}
    G_5 (z,\zP) = \left( kz \right)^{(d-3)/2} \left( k\zP \right)^{(d-3)/2} \, \tilde{G}_{5} ( z,\zP) \, ,
\end{equation}
such that $\tilde{G}_5$ satisfies
\begin{equation}
    \label{eq:A5-prop-simplified-diff-eqs}
     \left[ \partial_z^2 - \frac{\alpha(\alpha+1)}{z^2}  - \frac{p^2}{\xi} \right] \xi \tilde{G}_5 (z,\zP) = \delta(z-\zP),
\end{equation}
where we have defined $\alpha = -\frac{d-3}{2}$. With this, the propagator BCs become:
\begin{gather}
(1-P_{UV}) \tilde{G}_5^< (1/k,\zP) = 0, \quad P_{UV}\left( \partial_z +\frac{\alpha}{z} \right) \tilde{G}_5^< (z,\zP) \Big|_{z=1/k} = 0, \notag \\ \label{eq:BCs-A5-prop}
(1-P_{IR})\tilde{G}_5^> (1/T,\zP) = 0, \quad  P_{IR}\left( \partial_z +\frac{\alpha}{z}\right) \tilde{G}_5^> (z,\zP) \Big|_{z=1/T} = 0.
\end{gather}
Comparing this to the differential equation and BCs of the gauge bosons in Appendix~\ref{sec:prop_gauge_boson}, we can immediately write down the solution
\begin{equation}
    \tilde{G}_5(p,z,\zP) = \frac{1}{\xi} \tilde{G}_1\left(\frac{p}{\sqrt{\xi}},z,\zP\right), \ \, \text{with } \alpha\to-\frac{d-3}{2}, P_{UV} \leftrightarrow (1-P_{UV}), P_{IR} \leftrightarrow (1-P_{IR}) \,,
\end{equation}
and the expression for $G_1(p/\sqrt{\xi},z,\zP)$ given in Appendix~\ref{sec:prop_gauge_boson}.

\bibliographystyle{JHEP} 
\bibliography{GHGUT_Running}

\end{document}